\documentclass[12pt]{article}
\usepackage{latexsym,amssymb,amsfonts,amsmath}
\usepackage{graphicx}
\usepackage{color}
\usepackage{slashed}

\newlength{\extraspace}
\newlength{\extraspaces}
\newcommand{\be}{\begin{equation}
\addtolength{\abovedisplayskip}{\extraspaces}
\addtolength{\belowdisplayskip}{\extraspaces}
\addtolength{\abovedisplayshortskip}{\extraspace}
\addtolength{\belowdisplayshortskip}{\extraspace}}
\newcommand{\ee}{\end{equation}}

\newcommand{\bea}{\begin{eqnarray}
\addtolength{\abovedisplayskip}{\extraspaces}
\addtolength{\belowdisplayskip}{\extraspaces}
\addtolength{\abovedisplayshortskip}{\extraspace}
\addtolength{\belowdisplayshortskip}{\extraspace}}
\newcommand{\eea}{\end{eqnarray}}

\DeclareMathOperator{\Tr}{Tr}

\DeclareMathOperator{\hc}{H.c.}

\begin{document}
	
\addtolength{\baselineskip}{.8mm}

{\thispagestyle{empty}

\mbox{}                \hfill Revised: March 2022\hspace{1cm}\\

\begin{center}
{\Large\bf Study of some local and global $U(1)$ axial condensates in QCD at finite temperature}\\
\vspace*{2.0cm}
{\large
Nicoletta Carabba\footnote{nicoletta.carabba@uni.lu}
}\\
\vspace*{0.5cm}{\normalsize
{\textit{Department of Physics and Materials Science, University of Luxembourg,\\
L-1511 Luxembourg, Grand Duchy of Luxembourg}}}\\
\vspace*{0.5cm}
{\large
Enrico Meggiolaro\footnote{enrico.meggiolaro@unipi.it}
}\\
\vspace*{0.5cm}{\normalsize
{\textit{Dipartimento di Fisica, Universit\`a di Pisa,
and INFN, Sezione di Pisa,\\
Largo Pontecorvo 3, I-56127 Pisa, Italy}}}\\
\vspace*{2cm}{\large \bf Abstract}
\end{center}

\noindent
The aim of this work is to study some possible local and global $U(1)$ axial condensates in the high-temperature chirally restored phase of QCD, by means of two nonperturbative analytical techniques: (i) by expressing the functional averages in terms of the spectral density of the Euclidean Dirac operator and (ii) by evaluating the functional integrals in the instanton-background approximation. In this way, besides proving that these condensates are indeed different from zero in the high-temperature regime, we shall also derive their asymptotic temperature dependence and compare it with that of the topological susceptibility.
}
\newpage

\section{Introduction}

It is well known that the vacuum state of Quantum Chromo-Dynamics (QCD) is characterized by certain nonvanishing \textit{condensates} which cannot be understood in the framework of perturbation theory. In the so-called \textit{chiral limit}, in which $N_f$ quark masses are sent to zero ($N_f=2$ and $N_f=3$ being the physically relevant cases), the QCD Lagrangian turns out to be symmetric under the chiral group $U(1)_V \otimes U(1)_A \otimes SU(N_f)_V \otimes SU(N_f)_A$. In the quantum theory, the subgroup $SU(N_f)_V \otimes SU(N_f)_A$ is spontaneously broken down to $SU(N_f)_V$ because of the condensation of quark-antiquark pairs, which gives rise to the so-called \textit{chiral condensate} $\Sigma \equiv -\langle \overline{q}q \rangle = -\sum_{f=1}^{N_f}\langle \overline{q}_f(x) q_f(x) \rangle$, where the brackets $\langle \ldots \rangle$ stand for the vacuum expectation value at zero temperature or, more generally, for the thermal average at a finite temperature $T$. On the other hand, the $U(1)_A$ (axial) symmetry is broken by the \textit{quantum anomaly} \cite{Weinberg1975}. At the quantum level, under $U(1)_A$ transformations the action acquires a contribution proportional to the so-called \textit{topological charge} $Q$: although $Q$ is the integral of a total divergence, it can be nonzero because of the existence of topologically nontrivial gauge configurations known as \textit{instantons}, which are Euclidean solutions of the classical equations of motion with finite action and integer topological charge \cite{BPST1975,tHooft1976}.

Moreover, it is also known (mainly by lattice simulations \cite{HotQCD}) that, at a certain (pseudo)critical temperature $T_c\approx 150$ MeV, QCD (in the chiral limit) undergoes a phase transition which restores the $SU(N_f)_V \otimes SU(N_f)_A$ symmetry: the chiral condensate $\Sigma$, which is just an order parameter of this symmetry, vanishes above $T_c$.\\
Instead, the fate of the $U(1)_A$ symmetry above the transition remains unclear. Although the quantum anomaly is present at any finite temperature [so that an \emph{exact} restoration of the $U(1)_A$ symmetry is, of course, out of question], above some temperature $T_{U(1)}$ its effects could become practically negligible: if so, the $U(1)_A$ symmetry would be \emph{approximately} restored.
Two different scenarios are possible \cite{PW1984,Shuryak1994}: (i) this \emph{approximate} restoration could occur well inside the quark-gluon plasma phase, i.e., at temperatures sensibly larger than $T_c$ ($T_{U(1)} \gg T_c$), or, vice versa, (ii) it could occur simultaneously to the chiral one at $T_c$ ($T_{U(1)} \simeq T_c$); and the nature of the chiral phase transition at $T_c$ crucially depends on which of these two scenarios is realized.
For example, in the case $N_f=2$, if the first scenario is realized ($T_{U(1)} \gg T_c$), then the chiral phase transition is expected to be of second order, belonging to the three-dimensional $O(4)$ universality class; if, instead, the second scenario is realized ($T_{U(1)} \simeq T_c$), then the chiral phase transition may be either of first order or of second order but belonging to a universality class different from $O(4)$.
However, which of these two scenarios is indeed realized is still an (important) open question.

Traditionally, this question has been investigated, at least in the case $N_f=2$, by studying (mainly by lattice simulations) the so-called \textit{chiral susceptibilities} \cite{DK1987,Shuryak1994}. For each meson channel $M$ ($\sigma$, $\delta$, $\pi$, and $\eta$), the chiral susceptibility $\chi_M$ is defined as the integral over four-space of the two-point correlation function of the corresponding interpolating operator $J_M(x)=\overline{q}(x)\Gamma_M q(x)$, for some proper matrix $\Gamma_M$ in Dirac and flavour space:
\begin{equation}\label{meson}
J_\sigma= \overline{q}q,\quad
J^a_\delta=\overline{q}\tau_aq,\quad
J_\eta=i\overline{q}\gamma_5q,\quad
J^a_\pi=i\overline{q}\gamma_5\tau_aq,
\end{equation}
where $\tau_a$, with $a=1,2,3$, are the Pauli matrices, normalized so as $\Tr(\tau_a\tau_b)=2\delta_{ab}$. That is,
\begin{equation}\label{chiralsusc}
\chi_M=\int d^4x\, \langle J_M(x)J^{\dagger}_M(0)\rangle=\frac{1}{V}\int d^4x\int d^4y\, \langle J_M(x)J^{\dagger}_M(y)\rangle,
\end{equation}
where the brackets at the right-hand side stand for the functional integration over the gauge field and the quark fields and translational invariance has been used.\\
The importance of these correlators lies in the fact that under $SU(2)_A$ and $U(1)_A$ transformations the meson channels are mixed as follows:
\begin{equation}\label{symmpict}
\begin{split}
\sigma\quad&\underset{U(1)_A}{\longleftrightarrow}\quad \eta\\
SU(2)_A&\updownarrow\quad\updownarrow SU(2)_A\\
\pi\quad&\underset{U(1)_A}{\longleftrightarrow} \quad\delta
\end{split}
\end{equation}
The restoration of a certain symmetry thus results in the degeneracy between the correlation functions of the channels that are mixed under that symmetry.
In particular, besides the chiral condensate $\Sigma$, also the differences $\chi_\pi-\chi_\sigma$ and $\chi_\delta-\chi_\eta$ can be regarded (in the chiral limit $m\equiv m_{u,d} \to 0$) as order parameters of the $SU(2)_A$ symmetry and must vanish above the critical temperature $T_c$, as confirmed by lattice simulations. On the other hand, $\chi_\pi-\chi_\delta$ and $\chi_\sigma-\chi_\eta$ behave as order parameters of the $U(1)_A$ symmetry.\\
Several lattice simulations, measuring these quantities (for the case $N_f=2$ and also for the more realistic case $N_f=2+1$, with $m\equiv m_{u,d} \to 0$ and $m_s \sim 100$ MeV), have been carried out, but the results achieved so far are not yet conclusive.
Most of the studies \cite{lat1997,lat1998,lat1999,lat2000,lat2000bis,lat2011,lat2012,lat2014,Ding2021} (using \emph{staggered quarks} or \emph{domain-wall quarks} on the lattice) find that the $U(1)_A$-breaking difference $\chi_\pi-\chi_\delta$ is still sensibly nonzero above the chiral transition (so favouring the first scenario that we have mentioned above), but some others \cite{Cossu2013,Aoki2012} (using the so-called \emph{overlap quarks} on the lattice) find that this quantity vanishes for $T \ge T_c$, so indicating an \emph{effective} restoration of the $U(1)_A$ symmetry already at $T_c$, at least at the level of the chiral susceptibilities for the meson channels \eqref{meson} (and so favouring the second scenario that we have mentioned above).\footnote{We point out that here (and also in the rest of the paper) we are using the word ``\emph{effective} restoration'' (which is a bit stronger that just ``\emph{approximate} restoration'') with exactly the same meaning that was used in Refs. \cite{Cossu2013,Aoki2012}. Even if, as we have already said, the $U(1)_A$ symmetry is always broken by the quantum anomaly, it may happen that \emph{certain} (but not \emph{all}) correlation functions, obtained by considering the expectation values of operators which are not invariant under a $U(1)_A$ transformation, are \emph{exactly} equal to zero (in the chiral limit $m\to 0$) above $T_c$: in this case, we say that this particular set of correlation functions manifest an \emph{effective} restoration of the $U(1)_A$ symmetry above $T_c$. For example, the usual chiral condensate $\Sigma \equiv -\langle \overline{q}q \rangle$, which is also an order parameter for the $U(1)_A$ symmetry [the operator $\overline{q}q$ not being invariant under a $U(1)_A$ transformation], vanishes exactly (in the chiral limit $m\to 0$) for $T\ge T_c$. It was argued in Refs. \cite{Cossu2013,Aoki2012} (using both analytical and numerical methods on the lattice) that also the chiral susceptibilities of the meson channels \eqref{meson} (that is to say, the mass spectrum of these meson channels) manifest such an \emph{effective} restoration of the $U(1)_A$ symmetry for $T\ge T_c$.}

The aim of this work is to study other \textit{local} and \textit{global} ``genuine'' $U(1)$ axial condensates in the high-temperature chirally restored phase of QCD, by means of two nonperturbative analytical techniques: (i) by expressing the functional averages $\langle\ldots\rangle$ in terms of the \textit{spectral density} of the Euclidean Dirac operator $i\slashed{D}$ and (ii) by evaluating the functional integrals in the \textit{instanton}-background approximation.\\
The \textit{local} $U(1)$ axial condensates that we shall first consider are functional averages $\mathcal{C}_{U(1)}\equiv\langle \mathcal{O}_{U(1)}(x)\rangle$ of \textit{local} $2N_f$-quark operators of the form
\begin{equation}\label{U1det}
\mathcal{O}_{U(1)}(x)\sim\det_{st}\left[\overline{q}_s(x)\left(\frac{1+\gamma_5}{2}\right)q_t(x)\right]+\hc,
\end{equation}
where $s,t \in \{1,\ldots,N_f\}$ are flavour indices and the Dirac indices (not explicitly shown) are contracted in each quark bilinear $\overline{q}_s(x) \left(\frac{1+\gamma_5}{2}\right) q_t(x)$, while the colour indices (also not explicitly shown) can be contracted in different possible ways, so to give a colour singlet (see below): these operators are invariant under the whole chiral group except for the $U(1)_A$ transformations, so that, differently from the above-mentioned chiral condensate $\Sigma$ and the quantity $\chi_\pi-\chi_\delta$, their functional averages $\mathcal{C}_{U(1)}$ are ``genuine'' order parameters for the $U(1)_A$ symmetry alone (for any number of flavours).
Operators of this kind were first introduced by Kobayashi and Maskawa in 1970 \cite{KM1970}, as an additional effective vertex in a generalized Nambu-Jona-Lasinio model, and by 't Hooft in 1976 \cite{tHooft1976}, as an effective quark interaction in the background gauge field of an instanton. (See also Ref. \cite{Kunihiro2009} for an interesting historical review on this subject.)
These ``genuine'' $U(1)$ axial condensates were then reconsidered in Ref. \cite{EM1994} (in the context of an effective chiral Lagrangian formulation) and also in Ref. \cite{Shuryak1994}.\\
In this work, we shall also consider \textit{global} $U(1)$ axial condensates, taking the functional average of \textit{multilocal} operators of the form (indicating with $\epsilon^{f_1\ldots f_{N_f}}$ the completely antisymmetric tensor in the flavour indices $f_1,\ldots,f_{N_f} \in \{1,\ldots,N_f\}$, with $\epsilon^{12\ldots N_f} = 1$)
\begin{equation}\label{multilocal-operator}
\mathcal{O}_{U(1)}(x_1,\ldots,x_{N_f})\sim \epsilon^{f_1\ldots f_{N_f}}\overline{q}_1(x_1)\left(\frac{1+\gamma_5}{2}\right)q_{f_1}(x_1)\ldots\overline{q}_{N_f}(x_{N_f})\left(\frac{1+\gamma_5}{2}\right)q_{f_{N_f}}(x_{N_f}) +\hc
\end{equation}
[i.e., performing a ``point splitting'' of the $N_f$ quark bilinears contained in the expression of the local operator $\mathcal{O}_{U(1)}(x)$] and then integrating over the four-space coordinates.
The main motivation for introducing these new quantities is that [differently from the local $U(1)$ axial condensates] they can be studied using the spectral-density technique, as we shall see in detail in the next section. Moreover, the global feature of these new $U(1)$ axial condensates renders them promising objects for future numerical studies on the lattice (probably better than their local counterparts, for which, on the contrary, a direct numerical determination on the lattice is expected to be highly problematic: see, e.g., Ref. \cite{DM1995}).

The paper is organized as follows.\\
In the Appendix, we shall review, for the benefit of the reader, the results obtained for the chiral susceptibilities by using the spectral-density technique and, in section 2, we shall apply the same technique to study also the above-mentioned global $U(1)$ axial condensates and their relations with the chiral susceptibilities, as well as with the so-called \textit{topological susceptibility} $\chi_{top} \equiv \langle Q^2 \rangle/V$.\\
In section 3, instead, we shall explicitly compute the local and global $U(1)$ axial condensates in the high-temperature phase, using the instanton-background approximation of the functional integrals. In this way, besides proving that these condensates are indeed different from zero in the high-temperature regime, we shall also derive their asymptotic temperature dependence and compare it with that of the topological susceptibility $\chi_{top}$.\\
Finally, in section 4 we shall conclude by briefly summarizing the results obtained in this paper and giving also some prospects for future studies.

\section{Local and global $U(1)$ axial condensates and their spectral-density analysis}\label{rhocondensate}

\subsection{$U(1)$ axial condensates in QCD with $N_f=2$ light flavours}

Let us start by considering the most simple case (and, presumably, also the most relevant one, from the physical point of view), that is, the case of two light flavours.
It is known (see the third Ref. \cite{EM1994} and also Appendix A in Ref. \cite{EM2011}, where also the case $N_f=3$ is considered) that the most general \emph{local} quark operator (without derivatives) which has the required chiral transformation properties mentioned in the Introduction [i.e., it is invariant under the whole chiral group except for the $U(1)_A$ transformations] and is colour singlet, Hermitian, and $P$-invariant, is the following four-quark local operator:\footnote{Of course, this \emph{local} operator (as well as the other \emph{local} and \emph{multilocal} operators [Eqs. \eqref{U1det}, \eqref{multilocal-operator}, and \eqref{multilocal-operator_Nf=2}] that we shall discuss in the following) should be properly renormalized in a given renormalization scheme. This is surely a fundamental (and quite delicate) question, which is, however, beyond the original explorative scope of this paper and will be addressed in future works: in this paper, therefore, we shall simply neglect the renormalization question [apart from some brief comment in section 3, when discussing the computation in the instanton-background approximation of the local and global $U(1)$ axial condensates].}
\begin{equation}\label{local-operator_Nf=2}
\mathcal{O}^{(N_f=2)}_{U(1)}(x;\,\kappa_1,\kappa_2) = F^{\alpha\gamma}_{\beta\delta}(\kappa_1,\kappa_2)\,\epsilon^{st}\,\overline{q}^{\alpha}_1(x)\left(\frac{1+\gamma_5}{2}\right)q^{\beta}_s(x)\,\overline{q}^{\gamma}_2(x)\left(\frac{1+\gamma_5}{2}\right)q^{\delta}_t(x) +\hc,
\end{equation}
where $s, t \in \{1,2\}$ are flavour indices and $\epsilon^{st} = -\epsilon^{ts}$, $\epsilon^{12} = 1$, the Dirac indices (not explicitly shown) are contracted in each of the two quark bilinears
$\overline{q}^{\alpha}_1(x)\left(\frac{1+\gamma_5}{2}\right)q^{\beta}_s(x)$ and
$\overline{q}^{\gamma}_2(x)\left(\frac{1+\gamma_5}{2}\right)q^{\delta}_t(x)$,
while the Greek letters $\alpha,\beta,\gamma,\delta \in \{1,\ldots,N_c\}$ are colour indices and the colour tensor
$F^{\alpha\gamma}_{\beta\delta}(\kappa_1,\kappa_2)$ is given by
\begin{equation}\label{F-tensor}
F^{\alpha\gamma}_{\beta\delta}(\kappa_1,\kappa_2) \equiv \kappa_1\delta^{\alpha}_{\beta}\delta^{\gamma}_{\delta}+\kappa_2\delta^{\alpha}_{\delta}\delta^{\gamma}_{\beta},
\end{equation}
where $\kappa_1$ and $\kappa_2$ are arbitrary real constants. [Let us observe that if one chooses, in particular, $\kappa_1 = N_c$ and $\kappa_2 = -1$, then the operator \eqref{local-operator_Nf=2} just becomes, up to a proportionality constant, the effective quark interaction in the background gauge field of an instanton, found by 't Hooft in Ref. \cite{tHooft1976}.]

Let us now try to investigate, by means of the spectral-density technique, the \textit{local} $U(1)$ axial condensates obtained by taking the functional averages of the operators \eqref{local-operator_Nf=2}, $\mathcal{C}_{U(1)}\equiv\langle \mathcal{O}_{U(1)}(x)\rangle$.
By integrating over the quark fields, one obtains the following result:
\begin{equation}\label{U1condensate_Nf=2}
\begin{split}
\mathcal{C}^{(N_f=2)}_{U(1)} &= \frac{1}{2} F^{\alpha\gamma}_{\beta\delta} \langle \Tr_D\mathcal{G}_A^{\beta\alpha}(x,x)\Tr_D\mathcal{G}_A^{\delta\gamma}(x,x) + \Tr_D[\gamma_5\mathcal{G}_A^{\beta\alpha}(x,x)]\Tr_D[\gamma_5\mathcal{G}_A^{\delta\gamma}(x,x)]\\
&+ \Tr_D[\mathcal{G}_A^{\beta\gamma}(x,x)\mathcal{G}_A^{\delta\alpha}(x,x)] + \Tr_D[\gamma_5\mathcal{G}_A^{\beta\gamma}(x,x)\gamma_5\mathcal{G}_A^{\delta\alpha}(x,x)] \rangle,
\end{split}
\end{equation}
where $\mathcal{G}_A$ is the quark propagator in the external gauge field $A_\mu$ [see Eq.~\eqref{prop} with $f=f'$] and $\Tr_D$ stands for the trace over the Dirac indices only. By virtue of translational invariance, we can rewrite the $U(1)$ axial condensate $\mathcal{C}_{U(1)}$ integrating Eq.~\eqref{U1condensate_Nf=2} over $\frac{1}{V}\int d^4x$. Let us consider, for example, the third term of Eq.~\eqref{U1condensate_Nf=2}:
\begin{equation*}
F^{\alpha\gamma}_{\beta\delta}
\frac{1}{2V}\langle \int d^4x\,\sum_k \frac{u^{\beta,i}_k(x) u^{\dagger\gamma,j}_k(x)}{m-i\lambda_k}\sum_{k'}\frac{u^{\delta,j}_{k'}(x) u^{\dagger\alpha,i}_{k'}(x)}{m-i\lambda_{k'}}\rangle,
\end{equation*}
where $i,j \in \{1,\ldots,4\}$ are Dirac indices.
It is evident that the local feature of this condensate prevents us from using the orthonormality relation \eqref{orthonorm} of the $u_k$'s, since the four Dirac eigenfunctions are all evaluated at the same space point. If we want to perform the integration over four-space and to rewrite this object in terms of the spectral density, we should have the Dirac eigenfunctions with contracted quark indices evaluated at different space points. In other words, we should introduce a \emph{point splitting} and consequently perform an additional four-space integration:
\begin{equation*}
F^{\alpha\gamma}_{\beta\delta}
\frac{1}{2V}\langle \int d^4x\int d^4y\,\sum_k \frac{u^{\beta,i}_k(x) u^{\dagger\gamma,j}_k(y)}{m-i\lambda_k}\sum_{k'}\frac{u^{\delta,j}_{k'}(y) u^{\dagger\alpha,i}_{k'}(x)}{m-i\lambda_{k'}}\rangle.
\end{equation*}
To this purpose, let us define the following \emph{multilocal} operator:
\begin{equation}\label{multilocal-operator_Nf=2}
\mathcal{O}^{(N_f=2)}_{U(1)}(x,y)\equiv\epsilon^{st}\,\overline{q}^{\alpha}_1(x)\left(\frac{1+\gamma_5}{2}\right)q^{\alpha}_s(x)\,\overline{q}^{\gamma}_2(y)\left(\frac{1+\gamma_5}{2}\right)q^{\gamma}_t(y) +\hc,
\end{equation}
where, in order to guarantee the gauge invariance of the multilocal operator, only the colour contraction generated by the first term in Eq.~\eqref{F-tensor} (with $\kappa_1=1$) has been retained. We then define the \emph{global} $U(1)$ axial condensate $\mathcal{D}_{U(1)}$ as
\begin{equation}\label{U1global_Nf=2}
\begin{split}
&\mathcal{D}^{(N_f=2)}_{U(1)} \equiv \frac{1}{V}\int d^4x\int d^4y\,\langle\mathcal{O}^{(N_f=2)}_{U(1)}(x,y)\rangle\\
&= \frac{1}{V} \langle \det_{st} \left[ \int d^4x\,\overline{q}^{\alpha}_s(x)\left(\frac{1+\gamma_5}{2}\right)q^{\alpha}_t(x) \right]
+ \det_{st} \left[ \int d^4x\,\overline{q}^{\alpha}_s(x)\left(\frac{1-\gamma_5}{2}\right)q^{\alpha}_t(x) \right] \rangle .
\end{split}
\end{equation}
The multilocal operator \eqref{multilocal-operator_Nf=2} transforms under the chiral group exactly as the local operator \eqref{local-operator_Nf=2}, so that also $\mathcal{D}^{(N_f=2)}_{U(1)}$ is a genuine order parameter of the $U(1)_A$ symmetry: moreover, its global feature renders it somewhat similar to the chiral susceptibilities.

It is now easy to see that, performing the functional integration over the quark fields in the expression \eqref{U1global_Nf=2} for the new global condensate $\mathcal{D}^{(N_f=2)}_{U(1)}$, Eq.~\eqref{U1condensate_Nf=2} is replaced by
\begin{equation}\label{U1global_Nf=2_bis}
\mathcal{D}^{(N_f=2)}_{U(1)}=\frac{1}{2V}\langle (\Tr_{DCx}\mathcal{G}_A)^2 + [\Tr_{DCx}(\gamma_5\mathcal{G}_A)]^2 + \Tr_{DCx}(\mathcal{G}_A^2) + \Tr_{DCx}[(\gamma_5\mathcal{G}_A)^2] \rangle,
\end{equation}
where we have used the compact notation $\Tr_{DCx}$ for the trace over the Dirac, colour and spatial indices. By comparing the four terms of Eq.~\eqref{U1global_Nf=2_bis} with the expressions \eqref{sigma_conn,disc} and \eqref{eta_conn,disc}, we find that
\begin{gather*}
\frac{1}{2V}\langle\int d^4x\int d^4y\,\Tr_{DC}\mathcal{G}_A(x,x)\Tr_{DC}\mathcal{G}_A(y,y)\rangle=\frac{\chi_{\sigma,disc}}{8},\\
\frac{1}{2V}\langle\int d^4x\int d^4y\,\Tr_{DC}[\gamma_5\mathcal{G}_A(x,x)]\Tr_{DC}[\gamma_5\mathcal{G}_A(y,y)]\rangle=-\frac{\chi_{\eta,disc}}{8},\\
\frac{1}{2V}\langle\int d^4x\int d^4y\,\Tr_{DC}[\mathcal{G}_A(x,y)\mathcal{G}_A(y,x)]\rangle=-\frac{\chi_{\sigma,conn}}{4},\\
\frac{1}{2V}\langle\int d^4x\int d^4y\,\Tr_{DC}[\gamma_5\mathcal{G}_A(x,y)\gamma_5\mathcal{G}_A(y,x)]\rangle=\frac{\chi_{\eta,conn}}{4},
\end{gather*}
so that, summing the four contributions, the following result is found:
\begin{equation}\label{U1global_Nf=2_final}
\mathcal{D}^{(N_f=2)}_{U(1)} = \frac{1}{4}\left(\chi_{\eta,conn} - \chi_{\sigma,conn}\right) + \frac{1}{8}\left(\chi_{\sigma,disc} - \chi_{\eta,disc}\right) .
\end{equation}
It follows, in particular, that, in the chirally restored phase ($T>T_c$, $m=0$) where Eq.~\eqref{SU2rest} holds, the global $U(1)$ axial condensate $\mathcal{D}_{U(1)}$ in two-flavour QCD turns out to be proportional to the $U(1)_A$-breaking difference $\chi_\pi-\chi_\delta$, and, thus, to $\displaystyle\lim_{m\to 0}\chi_{top}/m^2$:
\begin{equation}\label{result}
\mathcal{D}^{(N_f=2)}_{U(1)}(T>T_c) = \frac{1}{2}(\chi_\pi-\chi_\delta) =
2\lim_{m\to 0}\frac{\chi_{top}}{m^2}.
\end{equation}
Using Eq.~\eqref{U1order_Nf=2} for $\chi_\pi-\chi_\delta$, we can immediately express $\mathcal{D}^{(N_f=2)}_{U(1)}$ for $T>T_c$ in terms of the spectral density:
\begin{equation*}
\mathcal{D}^{(N_f=2)}_{U(1)}(T>T_c) = 4\lim_{m\to 0}\int_{0}^{\infty}d\lambda\,\frac{m^2\rho(\lambda,m)}{(m^2+\lambda^2)^2}.
\end{equation*}
From this expression, one immediately recognizes that the new global condensate $\mathcal{D}^{(N_f=2)}_{U(1)}$ gets (for $T>T_c$) a nonzero contribution from the so-called \emph{Dirac zero modes}, whose contribution to the spectral density is given by $ \rho(\lambda,m)\vert_0 = \mathcal{C}m^2\delta(\lambda)$ [see the Appendix, Eq. \eqref{deltaC}]; in fact, substituing $\rho(\lambda,m) = \rho(\lambda,m)\vert_0 + \ldots = \mathcal{C}m^2\delta(\lambda) + \ldots$, one finds that
\begin{equation}\label{globaldelta}
\mathcal{D}^{(N_f=2)}_{U(1)}(T>T_c) = 2\mathcal{C} + \ldots,
\end{equation}
where the coefficient $\mathcal{C}$ is given by Eq.~\eqref{deltaC}.

\subsection{$U(1)$ axial condensates in QCD with $N_f$ light flavours}

Let us now consider the case of an arbitrary number $N_f>2$ of light flavours. As is well known (see the Appendix), if $N_f>2$, the chiral susceptibilities no longer represent good candidates to assess the $U(1)_A$ breaking and their place is taken by the expectation value of proper $2N_f$-quark operators, such as the global $U(1)$ axial condensate $\mathcal{D}_{U(1)}$, defined by generalizing Eq.~\eqref{U1global_Nf=2} to the more general case with $N_f$ light flavours:
\begin{multline}\label{U1global}
\mathcal{D}_{U(1)} \equiv \frac{1}{V} \int d^4x_1\dots\int d^4x_{N_f}\\
\times \left[ \langle\epsilon^{f_1\ldots f_{N_f}}\,\overline{q}^{\alpha_1}_1(x_1)\left(\frac{1+\gamma_5}{2}\right)q^{\alpha_1}_{f_1}(x_1)\ldots\overline{q}^{\alpha_{N_f}}_{N_f}(x_{N_f})\left(\frac{1+\gamma_5}{2}\right)q^{\alpha_{N_f}}_{f_{N_f}}(x_{N_f})\rangle \right.\\
+ \left. \langle\epsilon^{f_1\ldots f_{N_f}}\,\overline{q}^{\alpha_1}_1(x_1)\left(\frac{1-\gamma_5}{2}\right)q^{\alpha_1}_{f_1}(x_1)\ldots\overline{q}^{\alpha_{N_f}}_{N_f}(x_{N_f})\left(\frac{1-\gamma_5}{2}\right)q^{\alpha_{N_f}}_{f_{N_f}}(x_{N_f})\rangle \right].
\end{multline}
By performing the functional integration over the quark fields, one obtains $N_f!$ possible different Wick contractions. As an example, let us write the global condensate in the case $N_f=3$:
\begin{eqnarray}\label{U1global_Nf=3}
\mathcal{D}^{(N_f=3)}_{U(1)} &=& -\frac{1}{4V}\langle (\Tr_{DCx}\mathcal{G}_A)^3 + 3\Tr_{DCx}\mathcal{G}_A[\Tr_{DCx}(\gamma_5\mathcal{G}_A)]^2 + 3\big\{\Tr_{DCx}\mathcal{G}_A\Tr_{DCx}(\mathcal{G}_A^2) \nonumber\\
&+& \Tr_{DCx}\mathcal{G}_A\Tr_{DCx}[(\gamma_5\mathcal{G}_A)^2] + 2\Tr_{DCx}(\gamma_5\mathcal{G}_A)\Tr_{DCx}(\gamma_5\mathcal{G}_A^2) \big\} \nonumber\\
&+& 2\big\{\Tr_{DCx}(\mathcal{G}_A^3) + 3\Tr_{DCx}[\mathcal{G}_A(\gamma_5\mathcal{G}_A)^2]\big\}\rangle.
\end{eqnarray}
In the following, we shall concentrate on the fifth term of this expression and generalize it to the case of an arbitrary number $N_f$ of light flavours. Indeed, among the $N_f!$ possible different Wick contractions in the general case, we will always find a term proportional to
\begin{equation}\label{termN}
\frac{1}{V}\langle\Tr_{DCx}(\gamma_5\mathcal{G}_A)\Tr_{DCx}(\gamma_5\mathcal{G}_A^{N_f-1})\rangle.
\end{equation}
By using the anticommutativity of the Euclidean Dirac operator $i\slashed{D}$ with $\gamma_5$, the orthonormality relation \eqref{orthonorm} for the eigenfunctions $u_k$, and the well-known \emph{Atiyah-Singer theorem} [see the analogous derivation of Eq. \eqref{etadisc} in the Appendix], the first trace can be expressed in terms of the topological charge $Q$ as
\begin{equation}\label{traccia1}
\Tr_{DCx}(\gamma_5\mathcal{G}_A) = \int d^4x\,\Tr_{DC} \left[ \sum_k \frac{\gamma_5 u_k(x)u_k^\dagger(x)}{m-i\lambda_k} \right] = -\frac{Q}{m}.
\end{equation}
On the other hand, let us evaluate the following trace:
\begin{equation*}
\Tr_{DCx}(\gamma_5\mathcal{G}_A^N) = \int d^4x_1\dots\int d^4 x_N\,\Tr_{DC} \left[ \sum_{k_1} \frac{\gamma_5u_{k_1}(x_1)u_{k_1}^\dagger(x_2)}{m-i\lambda_{k_1}}\prod_{i=2}^{N}\sum_{k_i} \frac{u_{k_i}(x_i)u_{k_i}^\dagger(x_{i+1})}{m-i\lambda_{k_i}} \right],
\end{equation*}
with the boundary condition $x_{N+1}=x_1$. By virtue of the orthonormality relation \eqref{orthonorm} for the eigenfunctions $u_k$, the integration over $x_2,\ldots,x_N$ imposes $k_i=k$ $\forall\, i$, and, as a consequence, we derive that
\begin{equation}\label{tracciaN}
\Tr_{DCx}(\gamma_5\mathcal{G}_A^N) = \int d^4x_1\,\Tr_{DC} \left[ \sum_k \frac{\gamma_5 u_k(x_1)u_k^\dagger(x_1)}{(m-i\lambda_k)^N} \right] = -\frac{Q}{m^N}.
\end{equation}
Finally, substituting Eqs.~\eqref{traccia1} and \eqref{tracciaN} with $N=N_f-1$ into Eq.~\eqref{termN}, we find that
\begin{equation}\label{termN1}
\frac{1}{V}\langle\Tr_{DCx}(\gamma_5\mathcal{G}_A)\Tr_{DCx}(\gamma_5\mathcal{G}_A^{N_f-1})\rangle = \frac{1}{V}\frac{\langle Q^2\rangle}{m^{N_f}}=\frac{\chi_{top}}{m^{N_f}}.
\end{equation}
We note that this derivation is valid at any temperature $T$ and at any value of the (common) quark mass $m$. At high temperatures, the so-called \emph{dilute instanton gas approximation} (DIGA) \cite{GPY1981} predicts that the topological susceptibility $\chi_{top}$ is different from zero and depends on the (common) quark mass precisely as $m^{N_f}$ in the small-$m$ limit (and the same prediction is also derived using chiral effective Lagrangian models above $T_c$ \cite{EM2019,BM2020}): therefore, the contribution \eqref{termN1} is expected to be different from zero in the chiral limit $m\to 0$, and we can, thus, argue that it is responsible for the $U(1)_A$ breaking of the global condensate $\mathcal{D}_{U(1)}$ at high temperature, similarly to $\chi_\pi-\chi_\delta$ in the case $N_f=2$. Indeed, we have already demonstrated that, for $N_f=2$, the whole global condensate $\mathcal{D}_{U(1)}$ becomes proportional to $\displaystyle\lim_{m\to 0}\chi_{top}/m^2$ as the chiral symmetry is restored for $T>T_c$ [see Eq.~\eqref{result}]. It might be that the same thing happens also in the case $N_f>2$, i.e., that at high temperatures ($T>T_c$) and in the chiral limit $m\to 0$ the whole global condensate $\mathcal{D}_{U(1)}$ becomes proportional to $\displaystyle\lim_{m\to 0}\chi_{top}/m^{N_f}$: in the next section, we shall find a confirmation of this guess.

\section{Instanton-background computation of the local and global $U(1)$ axial condensates}

In the previous section (and in the Appendix), the crucial role of Dirac operator's zero modes in the $U(1)_A$ breaking has been remarked. This can be realized also through the following reasoning. Let $\mathcal{O}_{2n}$ be a $2n$-quark operator: after integrating over the quark fields, its functional average can be expressed as
\begin{equation}\label{funcaverage2}
\langle\mathcal{O}_{2n}\rangle=\frac{1}{Z}\int DA\,e^{-S_G[A]}\,[\det(\slashed{D}[A]+m)]^{N_f}\mathcal{G}_A^n,
\end{equation}
where $\mathcal{G}_A^n$ is a simplified notation for the product of $n$ external gauge-field propagators of the form \eqref{prop}, with properly contracted indices. Let us now consider the contribution of gauge configurations with a \emph{single} zero mode $\Psi_0$, such that $[\det(\slashed{D}+m)]^{N_f}\simeq m^{N_f}(\overline{\det}\slashed{D})^{N_f}$, where $\overline{\det}$ is restricted to the nonzero modes, and $\mathcal{G}_A\simeq \Psi_0\Psi_0^\dagger/m$ (apart from terms which are regular in $m$). In the case of an \emph{instanton} ($I$), according to our conventions (see the Appendix), the only zero mode is \emph{right-handed} (i.e., $\gamma_5 \Psi_0 = -\Psi_0$) and the topological charge is $Q=1$: the contribution to the functional integral \eqref{funcaverage2} in the chiral limit $m\to 0$ is then
\begin{equation}\label{instcontr}
\langle\mathcal{O}_{2n}\rangle_I\sim\frac{1}{Z}\int_I DA\,e^{-S_G[A]}\,(\overline{\det}\slashed{D}[A])^{N_f} m^{N_f}\Big(\frac{\Psi_0\Psi_0^\dagger}{m}\Big)^n.
\end{equation}
Evidently, if $n=N_f$, this contribution can be nonvanishing in the chiral limit. In particular, the product of two quark bilinears $J$ of the type given in Eq.~\eqref{meson} has $n=2$ and can, thus, receive a nonzero contribution of this type in two-flavour QCD. The same holds for the $U(1)$ axial condensates in the case of an arbitrary number of flavours $N_f$, as we will explicitly derive in this section.

\subsection{Instanton-background computation at zero temperature}

Here, we first perform an explicit computation of the $U(1)$ axial condensates in the instanton-background approximation at zero temperature, in the case $N_f=2$ and for an arbitrary number of colours $N_c$. The general case of an arbitrary number of light flavours $N_f$ at a finite temperature $T$ will be considered in the next subsection.
In order to restrict the path integral to the contribution of the instanton (that we denote as $\langle\ldots\rangle_I$), the integration over the gauge configurations $A_\mu$ is traded for the one over the instanton parameters: its orientation into the gauge group $SU(N_c)$, its centre $x_0$, and its scale size $\rho$, that is,
\begin{equation}\label{instintegration}
\frac{1}{Z}\int DA\,e^{-S_G}[\det(\slashed{D}+m)]^{N_f} \ldots \Longrightarrow \int dn(\rho) \int d^4x_0 \int dU \ldots,
\end{equation}
where $dU$ is the Haar measure of integration over the $SU(N_c)$ ``rotations'' of the instanton and $dn(\rho)$ is the measure of integration over the instanton size, defined as \cite{tHooft1976,SVZ1980,KapustaBook}
\begin{equation}\label{rhomeasure}
dn(\rho)=\frac{d\rho}{\rho^5}\,d(\rho),
\end{equation}
where the ``instanton density'' $d(\rho)$, near the chiral limit and for equal quark masses, can be approximated as
\begin{equation}\label{instdensity}
d(\rho) \mathop\simeq_{m\to 0} (m\rho)^{N_f} d_0(\rho),\quad \textrm{with}\quad d_0(\rho) = C_{N_c,N_f}\Big(\frac{8\pi^2}{g^2(\rho)}\Big)^{2N_c}e^{-\frac{8\pi^2}{g^2(\rho)}},
\end{equation}
where $C_{N_c,N_f}$ is a constant depending on both $N_c$ and $N_f$ and $g(\rho)$ is the running coupling constant at the length scale $\rho$ [see Eq. \eqref{runningcoupling} below].

Let us first consider the generic four-quark correlation function and integrate over the quark fields:
\begin{equation}\label{corr}
\langle \overline{q}_a^{\alpha,i}(\omega)q_b^{\beta,j}(x)\overline{q}_c^{\gamma,k}(y)q_d^{\delta,l}(z)\rangle = \delta_{ab}\delta_{cd}\langle\mathcal{G}^{\beta\alpha,ji}_{A}(x,\omega)\mathcal{G}^{\delta\gamma,lk}_{A}(z,y)\rangle -\delta_{ad}\delta_{cb}\langle\mathcal{G}^{\delta\alpha,li}_{A}(z,\omega)\mathcal{G}^{\beta\gamma,jk}_{A}(x,y)\rangle,
\end{equation}
where $a,b,c$, and $d$ are flavour indices, while Greek letters $\alpha,\beta,\gamma$, and $\delta$ stand for colour indices, and $i,j,k$, and $l$ are Dirac indices.
In the instanton background, the Dirac operator has a zero mode of the form \cite{tHooft1976}
\begin{equation}\label{0modetotal}
\Psi^{\alpha,i}_0(x-x_0,\rho)=\psi_0(x-x_0,\rho)v_{\alpha,i},
\end{equation}
where the spinors $v_\alpha$ satisfy the following relation:
\begin{equation}\label{spinor-v}
\sum_\alpha v_{\alpha,i} v^\dagger_{\alpha,j} = \left(\frac{1-\gamma_5}{4}\right)_{ij}
\end{equation}
and the function $\psi_0(x-x_0,\rho)$ is given by
\begin{equation}\label{0mode}
\psi_0(x-x_0,\rho) = \frac{\sqrt{2}}{\pi}\frac{\rho}{[\rho^2+(x-x_0)^2]^{3/2}},
\end{equation}
so that the following identities hold:
\begin{equation}\label{normpsi0}
\int d^4x\,[\psi_0(x-x_0,\rho)]^2=1,\qquad \int d^4x\,[\psi_0(x-x_0,\rho)]^4=\frac{1}{5\pi^2\rho^4}.
\end{equation}
Near the chiral limit ($m\to 0$), the quark propagator for each flavour [see Eq. \eqref{prop} with $f=f'$] can be expressed as
\begin{equation}\label{instprop}
\mathcal{G}^{\alpha\beta,ij}_{A}(x,y) = \sum_k\frac{u^{\alpha,i}_k(x)u^{\dagger\beta,j}_k(y)}{m-i\lambda_k} \simeq \frac{\Psi^{\alpha,i}_0(x-x_0,\rho)\Psi^{\dagger\beta,j}_0(y-x_0,\rho)}{m},
\end{equation}
where the regular terms in $m$ have been neglected. Substituting into Eq.~\eqref{corr}, and making use of Eq. \eqref{instintegration}, one obtains the following expression for the four-quark correlation function in the instanton background:
\begin{equation}\label{corr2}
\begin{split}
&\langle \overline{q}_a^{\alpha,i}(\omega)q_b^{\beta,j}(x)\overline{q}_c^{\gamma,k}(y)q_d^{\delta,l}(z)\rangle_I
= (\delta_{ab}\delta_{cd}-\delta_{ad}\delta_{cb}) \langle v_{\beta,j}v^\dagger_{\alpha,i}v_{\delta,l}v^\dagger_{\gamma,k} \rangle_{SU(N_c)}\\
&\times \int\frac{dn(\rho)}{m^2} \int d^4x_0\, \psi_0(x-x_0,\rho)\psi_0(\omega-x_0,\rho)\psi_0(z-x_0,\rho)\psi_0(y-x_0,\rho),
\end{split}
\end{equation}
where $\langle\ldots\rangle_{SU(N_c)}$ is the average over the possible $SU(N_c)$ ``rotations'' of the spinor $v$ (and $v^\dagger$), i.e.,
\begin{equation}\label{groupint}
\langle v_{\beta}v^\dagger_{\alpha}v_{\delta}v^\dagger_{\gamma} \rangle_{SU(N_c)} \equiv \int dU v'_\beta v'^\dagger_{\alpha} v'_{\delta} v'^\dagger_{\gamma},\quad \textrm{where}\quad
v'_\alpha = U_{\alpha\alpha'}v_{\alpha'},\quad v'^\dagger_\alpha = v^\dagger_{\alpha'} U^\dagger_{\alpha'\alpha} ,
\end{equation}
and $dU$ is the Haar invariant measure over the group $SU(N_c)$.
Using the following rule of integration over $SU(N_c)$ \cite{CreutzBook}:
\begin{eqnarray*}
\lefteqn{
\int dU\, U_{\beta\beta'}U^\dagger_{\alpha'\alpha}U_{\delta\delta'}U^\dagger_{\gamma'\gamma} = \frac{1}{N_c^2-1}(\delta_{\beta\alpha}\delta_{\beta'\alpha'}\delta_{\delta\gamma}\delta_{\delta'\gamma'} + \delta_{\beta\gamma}\delta_{\beta'\gamma'}\delta_{\delta\alpha}\delta_{\delta'\alpha'}) }\\
& & - \frac{1}{N_c(N_c^2-1)}(\delta_{\beta\alpha}\delta_{\beta'\gamma'}\delta_{\delta\gamma}\delta_{\delta'\alpha'}+\delta_{\beta\gamma}\delta_{\beta'\alpha'}\delta_{\delta\alpha}\delta_{\delta'\gamma'}),
\end{eqnarray*}
and the relation \eqref{spinor-v}, one obtains
\begin{equation*}
\langle v_{\beta,j}v^\dagger_{\alpha,i}v_{\delta,l}v^\dagger_{\gamma,k} \rangle_{SU(N_c)} = \frac{1}{4N_c(N_c^2-1)}\Big[(N_c\delta_{\beta\alpha}\delta_{\delta\gamma}-1\delta_{\beta\gamma}\delta_{\delta\alpha})\Big(\frac{1-\gamma_5}{2}\Big)_{ji}\Big(\frac{1-\gamma_5}{2}\Big)_{lk}+\binom{\alpha\leftrightarrow\gamma}{i\leftrightarrow k}\Big].
\end{equation*}
Finally, substituting this expression into Eq.~\eqref{corr2}, one finds the following result:
\begin{equation}\label{corr3}
\begin{split}
&\langle \overline{q}_a^{\alpha,i}(\omega)q_b^{\beta,j}(x)\overline{q}_c^{\gamma,k}(y)q_d^{\delta,l}(z)\rangle_I = (\delta_{ab}\delta_{cd}-\delta_{ad}\delta_{cb})\\
&\times \frac{1}{4N_c(N_c^2-1)}\Big[(N_c\delta_{\beta\alpha}\delta_{\delta\gamma}-\delta_{\beta\gamma}\delta_{\delta\alpha})\Big(\frac{1-\gamma_5}{2}\Big)_{ji}\Big(\frac{1-\gamma_5}{2}\Big)_{lk}+\binom{\alpha\leftrightarrow\gamma}{i\leftrightarrow k}\Big]\\
&\times \int\frac{dn(\rho)}{m^2} \int d^4x_0\, \psi_0(x-x_0,\rho)\psi_0(\omega-x_0,\rho)\psi_0(z-x_0,\rho)\psi_0(y-x_0,\rho).
\end{split}
\end{equation}
This expression is valid for any number $N_c$ of colours, generalizing a relation found by Callan, Dashen, and Gross in Ref. \cite{CDG1978}. (Our result has an extra factor $1/4$, due to the fact that the zero mode used in Refs. \cite{tHooft1976,CDG1978} was erroneously normalized to $2$, as noted also in Ref. \cite{SVZ1980}, instead of $1$ as in our case.) Moreover, Eq.~\eqref{corr3} is valid for any number $N_f$ of flavours, although in what follows we shall use it (at first) in the particular case $N_f=2$, since only in this case is $\mathcal{O}_{U(1)}$ a four-quark operator.

\subsubsection{Local $U(1)$ axial condensate (in the case $N_f=2$)}

In the \emph{local} case $x=y=z=\omega$, the integration over the instanton centre $x_0$ in Eq. \eqref{corr3} can be immediately performed, making use of the second Eq.~\eqref{normpsi0}, and the result does not depend on $x$, as expected from translational invariance:
\begin{equation}\label{corrlocal}
\begin{split}
&\langle \overline{q}_a^{\alpha,i}(x)q_b^{\beta,j}(x)\overline{q}_c^{\gamma,k}(x)q_d^{\delta,l}(x)\rangle_I = (\delta_{ab}\delta_{cd}-\delta_{ad}\delta_{cb}) \int\frac{dn(\rho)}{5\pi^2 m^2\rho^4}\\
&\times \frac{1}{4N_c(N_c^2-1)}\Big[(N_c\delta_{\beta\alpha}\delta_{\delta\gamma}-\delta_{\beta\gamma}\delta_{\delta\alpha})\Big(\frac{1-\gamma_5}{2}\Big)_{ji}\Big(\frac{1-\gamma_5}{2}\Big)_{lk}+\binom{\alpha\leftrightarrow\gamma}{i\leftrightarrow k}\Big].\\
\end{split}
\end{equation}
Using this result, we can compute the local $U(1)$ axial condensate $\mathcal{C}_{U(1)}\vert_I \equiv \langle\mathcal{O}_{U(1)}\rangle_I$ in the instanton background, in the case $N_f=2$. The four-quark operator $\mathcal{O}^{(N_f=2)}_{U(1)}$ is defined by Eq.~\eqref{local-operator_Nf=2}, which, using the properties $F^{\alpha\gamma}_{\beta\delta} = F^{\gamma\alpha}_{\delta\beta}$ and $F^{\alpha\gamma}_{\beta\delta} = F^{\beta\delta}_{\alpha\gamma}$ of the colour tensor \eqref{F-tensor}, can be expressed as
\begin{equation}\label{operatorcorr}
\begin{split}
&\mathcal{O}^{(N_f=2)}_{U(1)}(x) = F^{\alpha\gamma}_{\beta\delta}\epsilon^{st}
\overline{q}_1^{\alpha,i}(x)q_s^{\beta,j}(x)\overline{q}_2^{\gamma,k}(x)q_t^{\delta,l}(x)\\
& \times \left[\Big(\frac{1+\gamma_5}{2}\Big)_{ij}\Big(\frac{1+\gamma_5}{2}\Big)_{kl} + \Big(\frac{1-\gamma_5}{2}\Big)_{ij}\Big(\frac{1-\gamma_5}{2}\Big)_{kl}\right].
\end{split}
\end{equation}
Thus, to compute $\langle O^{(N_f=2)}_{U(1)} \rangle_I$, we just need to consider Eq.~\eqref{corrlocal} with $a=1$, $b=s$, $c=2$, and $d=t$: the contraction over the flavour indices yields a factor $(\delta_{1s}\delta_{2t}-\delta_{1t}\delta_{2s})\epsilon^{st}=2$.\\
Contracting the colour tensor \eqref{F-tensor} with the two different colour factors appearing in the two terms on the right-hand side of Eq. \eqref{corrlocal}, one obtains
$F^{\alpha\gamma}_{\beta\delta} (N_c\delta_{\beta\alpha}\delta_{\delta\gamma}-\delta_{\beta\gamma}\delta_{\delta\alpha}) = \kappa_1 N_c(N_c^2-1)$
in the first term and
$F^{\alpha\gamma}_{\beta\delta} (N_c\delta_{\beta\gamma}\delta_{\delta\alpha}-\delta_{\beta\alpha}\delta_{\delta\gamma}) = \kappa_2 N_c(N_c^2-1)$
in the second one.\\
Finally, since the products of different chiral projectors vanish, when one takes the trace over the Dirac indices, the term in Eq.~\eqref{operatorcorr} containing the left projectors $\big(\frac{1+\gamma_5}{2}\big)_{ij}\big(\frac{1+\gamma_5}{2}\big)_{kl}$ does not contribute to $\langle \mathcal{O}_{U(1)} \rangle_I$, and one is, thus, left with these two Dirac contractions:
\begin{equation*}
\Big(\frac{1-\gamma_5}{2}\Big)_{ij}\Big(\frac{1-\gamma_5}{2}\Big)_{kl}\Big(\frac{1-\gamma_5}{2}\Big)_{ji}\Big(\frac{1-\gamma_5}{2}\Big)_{lk}=\Big[\Tr\Big(\frac{1-\gamma_5}{2}\Big)\Big]^2=4,
\end{equation*}
\begin{equation*}
\Big(\frac{1-\gamma_5}{2}\Big)_{ij}\Big(\frac{1-\gamma_5}{2}\Big)_{kl}\Big(\frac{1-\gamma_5}{2}\Big)_{jk}\Big(\frac{1-\gamma_5}{2}\Big)_{li}=\Tr\Big(\frac{1-\gamma_5}{2}\Big)=2.
\end{equation*}
The computation can be repeated for an \emph{anti-instanton} ($\bar{I}$) configuration (for which $Q=-1$) just by replacing $\gamma_5$ with $-\gamma_5$. In both cases, one thus obtains the same result:
\begin{equation}\label{instcondensato}
\mathcal{C}^{(N_f=2)}_{U(1)}\vert_I = \mathcal{C}^{(N_f=2)}_{U(1)}\vert_{\bar{I}} = (2\kappa_1+\kappa_2)\int \frac{dn(\rho)}{5\pi^2 m^2\rho^4}.
\end{equation}
One can also sum the two contributions and define
$\mathcal{C}^{(N_f=2)}_{U(1)}\vert_{inst} \equiv \mathcal{C}^{(N_f=2)}_{U(1)}\vert_I + \mathcal{C}^{(N_f=2)}_{U(1)}\vert_{\bar{I}} = 2\mathcal{C}^{(N_2=2)}_{U(1)}\vert_I$
as the contribution resulting from \emph{both} the instanton (for which $Q=1$) and the anti-instanton (for which $Q=-1$): using Eqs.~\eqref{rhomeasure} and \eqref{instdensity} with $N_f=2$, it is therefore given by
\begin{equation}\label{instcondensato2}
\mathcal{C}^{(N_f=2)}_{U(1)}\vert_{inst} \mathop\simeq_{m\to 0} 2(2\kappa_1+\kappa_2)\int_{0}^{\infty} \frac{d\rho}{5\pi^2\rho^7}\,d_0(\rho) = \frac{2(2\kappa_1+\kappa_2)C_{N_c,N_f}}{5\pi^2}\int_{0}^{\infty} \frac{d\rho}{\rho^7}\,\Big(\frac{8\pi^2}{g^2(\rho)}\Big)^{2N_c}e^{-\frac{8\pi^2}{g^2(\rho)}},
\end{equation}
where we notice that the dependence on $m$ has cancelled out, so that the result is nonzero in the chiral limit $m\to 0$.
We observe that this integral is, however, affected by an infrared divergence as $\rho\to\infty$. Indeed, substituting the one-loop result for the running coupling constant,
\begin{equation}\label{runningcoupling}
g^2(\rho)=\frac{1}{2\beta_0\ln(\frac{1}{\rho\Lambda_{QCD}})},\quad \textrm{with}\quad \beta_0=\frac{1}{(4\pi)^2}\left(\frac{11}{3}N_c-\frac{2}{3}N_f\right),
\end{equation}
one finds that
$e^{-\frac{8\pi^2}{g^2(\rho)}}\simeq (\rho\Lambda_{QCD})^{b_0}$, with $b_0=(4\pi)^2\beta_0=\frac{11}{3}N_c-\frac{2}{3}N_f$:
for $N_f=2$ and $N_c=3$, $b_0 \simeq 9.6$ and, therefore, $\mathcal{C}_{U(1)}\vert_{inst}$ diverges as $\int_{0}^{\infty} d\rho\, \rho^{b_0-7}$.\\
However, we note that Eq.~\eqref{runningcoupling} is valid only for $\rho\ll1/\Lambda_{QCD}$: for larger values of the (anti-)instanton size $\rho$, one cannot rely on the perturbative expression for the running coupling constant $g(\rho)$. As is well known (and as we shall see in the next subsection), this infrared problem disappears when one considers the theory at a finite temperature $T$, $1/T$ acting in this case as an infrared cutoff.

\subsubsection{Global $U(1)$ axial condensate (in the case $N_f=2$)}

Using Eq.~\eqref{corr3} in the case $x=\omega$ and $y=z$, we can also compute the \emph{global} $U(1)$ axial condensate $\mathcal{D}_{U(1)}$ defined by Eq.~\eqref{U1global_Nf=2} [see also the expression \eqref{U1global}] in the particular case $N_f=2$.
The only substantial difference with respect to the computation in the local case is the integration over the instanton centre $x_0$, which in this case gives
\begin{equation}\label{0modeglobal}
\frac{1}{V} \int d^4x_0 \int d^4x\,[\psi_0(x-x_0,\rho)]^2 \int d^4y\,[\psi_0(y-x_0,\rho)]^2 = 1,
\end{equation}
by virtue of the first Eq. \eqref{normpsi0}. Adjusting properly the passages above (taking $\kappa_1=1$ and $\kappa_2=0$) and using Eqs.~\eqref{rhomeasure} and \eqref{instdensity} with $N_f=2$, one immediately finds that
\begin{equation}\label{instglobal}
\mathcal{D}^{(N_f=2)}_{U(1)}\vert_{inst} \equiv 2\mathcal{D}^{(N_f=2)}_{U(1)}\vert_I = 4\int \frac{dn(\rho)}{m^2} \mathop\simeq_{m\to 0} 4\int_0^\infty \frac{d\rho}{\rho^3}\, d_0(\rho) ,
\end{equation}
i.e., a nonzero result in the chiral limit $m\to 0$.

The result \eqref{instglobal} is consistent with the one found in Eq. \eqref{globaldelta} by considering the contribution of the zero modes to the spectral density, taking into account that [see Eq. \eqref{deltaC} in the case $N_f=2$] $\mathcal{C} = \frac{1}{V}\frac{\langle n_0 \rangle}{m^2}$, where now $\langle n_0 \rangle \simeq \langle n_0 \rangle_I + \langle n_0 \rangle_{\bar{I}} = 2V\int dn(\rho)$.
We recall that Eq.~\eqref{globaldelta} was derived above the chiral transition, and indeed we also expect the instanton-background approximation to be reliable only above the chiral transition: therefore, in the above-reported expressions, $dn(\rho)$ should be more properly replaced by its expression at a finite temperature $T$, which will be considered and discussed in the next subsection.

\subsection{Instanton-background computation at high temperatures $T$}

In Ref. \cite{GPY1981}, Gross, Pisarski, and Yaffe demonstrated that at a finite temperature $T$ instantons of size $\rho\geq \frac{1}{T}$ are suppressed, rendering integrals such as \eqref{instcondensato} convergent in the infrared and dominated by $\rho\sim\frac{1}{T}$: it follows that as $T\to\infty$ all the instantons are suppressed. At high enough temperatures, one can use perturbation theory, being $\rho\ll1$ and, thus, $g(\rho)\ll1$ for the relevant configurations.

Performing a one-loop computation in the instanton-background approximation, Gross, Pisarski, and Yaffe found the following expression for the ``reduced instanton density'' $d_0(\rho,T)$ at a finite temperature $T$:
\begin{equation}\label{reduceddensityT}
d_0(\rho,T)=C_{N_c,N_f}\Big(\frac{8\pi^2}{g^2(\rho)}\Big)^{2N_c}\exp\Big\{-\Big[\frac{8\pi^2}{g^2(\rho)}+\lambda^2\Big(\frac{2N_c+N_f}{3}\Big)+A(\lambda)\Big]\Big\},
\end{equation}
where $\lambda\equiv\pi\rho T$ and $A(\lambda)$ diverges logarithmically as $\lambda\to\infty$ [if we are interested in only the asymptotic temperature dependence, the explicit form of $A(\lambda)$ is irrelevant].\\
Besides the instanton density, also the zero mode \eqref{0mode} has to be modified at a finite temperature $T$ (see below): however, if one is considering the global $U(1)$ axial condensate $\mathcal{D}_{U(1)}$ or also the chiral susceptibilities, its exact expression is irrelevant, since these quantities depend on the zero mode only through its normalization condition [see Eq.~\eqref{0modeglobal} and the first Eq. \eqref{normpsi0}], which is valid at any temperature $T$.
In fact, being $\gamma_5 \Psi_0 = \mp \Psi_0$, and thus $\gamma_5 \mathcal{G}_A = \mp \mathcal{G}_A$ [see Eq. \eqref{instprop}], in the field of an instanton or anti-instanton, the explicit expressions \eqref{U1global_Nf=2_bis} for $\mathcal{D}^{(N_f=2)}_{U(1)}$ and \eqref{U1global_Nf=3} for $\mathcal{D}^{(N_f=3)}_{U(1)}$, that we have found in the previous section, reduce (when evaluated in the instanton-background approximation) to the functional averages of various terms containing products of factors of the type $\Tr_{DCx}(\mathcal{G}_A^N)$, which, by virtue of Eq. \eqref{instprop} and the normalization condition $\int d^4x\, \Psi_0^{\dagger\alpha,i}(x-x_0,\rho) \Psi_0^{\alpha,i}(x-x_0,\rho) = 1$ [see Eq. \eqref{orthonorm}], is equal to $1/m^N$.

\subsubsection{Global $U(1)$ axial condensate and chiral susceptibilities}

Therefore, by virtue of the above-reported considerations, the expression \eqref{instglobal} for the global $U(1)$ axial condensate in the case $N_f=2$ is generalized as follows, at a finite temperature $T$:
\begin{equation}\label{highTglobal}
\begin{split}
\mathcal{D}^{(N_f=2)}_{U(1)}(T)\vert_{inst} &\mathop\simeq_{m\to 0} 4\int_{0}^{\infty} \frac{d\rho}{\rho^3}\, d_0(\rho,T)\sim \int_{0}^{\infty} \frac{d\rho}{\rho^3}\, (\rho\Lambda_{QCD})^{b_0}e^{-\frac{2}{3}(\pi\rho T)^2(N_c+1)}\\
&= \Lambda^{b_0}_{QCD} (\pi T)^{2-b_0} \int_{0}^{\infty} d\lambda\, \lambda^{b_0-3}e^{-\frac{2}{3}\lambda^2(N_c+1)}.
\end{split}
\end{equation}
That is to say, we have found the following asymptotic temperature dependence of $\mathcal{D}_{U(1)}$ in the case $N_f=2$:
\begin{equation}\label{U1global_Nf=2_T}
\mathcal{D}^{(N_f=2)}_{U(1)}(T) \sim T^{2-b_0},
\end{equation}
where $b_0=(11N_c-2N_f)/3=(11N_c-4)/3$. [Let us observe that, for $N_f=2$, $\mathcal{D}_{U(1)}$ has dimension $2$ in natural units, and indeed the expression above must be intended to be multiplied by $\Lambda^{b_0}_{QCD}$, as explicitly shown in Eq.~\eqref{highTglobal}.]\\
Now, recalling that for $T>T_c$ (in the chiral limit $m\to 0$) $\mathcal{D}_{U(1)}$ comes out to be proportional to $\chi_{\pi}-\chi_{\delta}$ and to $\displaystyle\lim_{m\to 0}\chi_{top}/m^2$ [see Eq.~\eqref{result}], we also obtain that
\begin{equation}\label{suscT_Nf=2}
(\chi_\pi-\chi_\delta)(T)\vert_{N_f=2} \sim T^{2-b_0},
\end{equation}
and, for the topological susceptibility,
\begin{equation}\label{chitopT_Nf=2}
\chi_{top}(T)\vert_{N_f=2} \propto m^2\mathcal{D}^{(N_f=2)}_{U(1)}(T) \sim T^{4-b_0}\Big(\frac{m}{T}\Big)^2.
\end{equation}
which coincides with the well-known DIGA prediction for the asymptotic temperature dependence of $\chi_{top}$ \cite{GPY1981} in the case of $N_f=2$ light flavours. [This expression is usually derived directly from the free energy $F(\theta,T)$, being $\chi_{top}=\frac{\partial^2F}{\partial \theta^2}\vert_{\theta=0}$.] Moreover, let us note the correctness of the dimensions also in Eqs. \eqref{suscT_Nf=2} and \eqref{chitopT_Nf=2}, the chiral and topological susceptibilities having dimensions $2$ and $4$, respectively.

It is easy to generalize the asymptotic behaviours \eqref{U1global_Nf=2_T}--\eqref{chitopT_Nf=2} to the case of an arbitrary number of light flavours $N_f$. Indeed, as already pointed out, the result \eqref{corr3} is valid also in this more general case, and, therefore, at high $T$, the chiral susceptibilities (being expectation values of certain four-quark operators) are always proportional to the integral in Eq.~\eqref{instglobal}, with $dn(\rho,T) = \frac{d\rho}{\rho^5}\,d(\rho,T)$ and $d(\rho,T ) \simeq (m\rho)^{N_f}d_0(\rho,T)$:
\begin{equation}\label{suscT}
\begin{split}
(\chi_\pi-\chi_\delta)(T)\vert_{inst} &\sim \int \frac{dn(\rho,T)}{m^2} \sim m^{N_f-2}\int_{0}^{\infty} \frac{d\rho}{\rho^{5-N_f}}\, (\rho\Lambda_{QCD})^{b_0}e^{-\frac{2N_c+N_f}{3}(\pi\rho T)^2} \\
&\sim T^{2-b_0}\Big(\frac{m}{T}\Big)^{N_f-2}.
\end{split}
\end{equation}
The contribution of the instantons to the chiral susceptibilities thus vanishes, in the case $N_f>2$, as $m^{N_f-2}$ in the chiral limit $m\to 0$, just as the contribution of the Dirac zero modes [which give rise to a term $\rho(\lambda,m)\vert_0 = \mathcal{C} m^{N_f}\delta(\lambda)$ in the spectral density; see the Appendix].
For a comparison, the contribution of the instantons to the chiral condensate $\Sigma \equiv -\langle \overline{q}q \rangle$ at high temperatures $T$ turns out to be
\be
\Sigma\vert_{inst} = \frac{2N_f}{V} \langle \Tr_{DCx} \mathcal{G_A} \rangle_I = 2N_f \int \frac{dn(\rho,T)}{m} \mathop\simeq_{m\to 0} 2N_f m^{N_f-1} \int_{0}^{\infty}\frac{d\rho}{\rho^{5-N_f}}\, d_0(\rho,T),
\ee
which correctly vanishes in the chiral limit $m\to 0$ for every $N_f \ge 2$.

Finally, let us consider the global $U(1)$ axial condensate $\mathcal{D}_{U(1)}$ for an arbitrary $N_f$, defined by Eq.~\eqref{U1global}: this condensate being the expectation value of a certain $2N_f$-quark operator, it will \emph{not} be proportional (as the chiral susceptibilities) to the integral $\int dn(\rho,T)/m^2$ but rather to the integral $\int dn(\rho,T)/m^{N_f}$, where the factor $1/m^{N_f}$ comes from the $N_f$ propagators of the form \eqref{instprop}. We thus obtain the following asymptotic behaviour:
\begin{equation}\label{U1global_T}
\mathcal{D}_{U(1)}(T)\vert_{inst} \sim \int \frac{dn(\rho,T)}{m^{N_f}} \mathop\simeq_{m\to 0} \int_{0}^{\infty}\frac{d\rho}{\rho^{5-N_f}}\, d_0(\rho,T) \sim T^{4-N_f-b_0}.
\end{equation}
[Let us observe that the dimension of $\mathcal{D}_{U(1)}$ is $3N_f-4(N_f-1)=4-N_f$.]\\
We have, thus, found a nonzero result for the global condensate $\mathcal{D}_{U(1)}$ at high temperatures and in the chiral limit $m\to 0$:
consistently with the fact that the instantons are asymptotically suppressed at high temperatures, the $U(1)$ axial condensate $\mathcal{D}_{U(1)}$ vanishes as $T\to\infty$, but it is anyhow different from zero at any finite temperature.\\
At the end of the previous section, we have demonstrated that $\mathcal{D}_{U(1)}$ contains a term [see Eq.~\eqref{termN1}] proportional to $\chi_{top}/m^{N_f}$, and we have guessed that the whole global condensate $\mathcal{D}_{U(1)}$ might be proportional to $\displaystyle\lim_{m\to 0}\chi_{top}/m^{N_f}$ above the chiral phase transition, in the chiral limit $m\to 0$.
Indeed, the result \eqref{U1global_T} is consistent with this guess and with the above-mentioned DIGA prediction for $\chi_{top}$ \cite{GPY1981}, i.e.,
\begin{equation}\label{chitopT}
\chi_{top}(T) \propto m^{N_f}\mathcal{D}_{U(1)}(T) \sim T^{4-b_0}\Big(\frac{m}{T}\Big)^{N_f}.
\end{equation}
An even more direct confirmation of this result can also be found by evaluating, in the instanton-background approximation, the explicit expressions \eqref{U1global_Nf=2_bis} for $\mathcal{D}^{(N_f=2)}_{U(1)}$ and \eqref{U1global_Nf=3} for $\mathcal{D}^{(N_f=3)}_{U(1)}$ that we have found in the previous section.
Since in the field of an instanton or anti-instanton one has that $\gamma_5 \mathcal{G}_A = \mp \mathcal{G}_A$ [see Eq. \eqref{instprop}, being $\gamma_5 \Psi_0 = \mp \Psi_0$] and $Q = \pm 1$, one can easily compute the expressions \eqref{U1global_Nf=2_bis} and \eqref{U1global_Nf=3} in the instanton-background approximation, making use of the relations \eqref{traccia1} and \eqref{tracciaN}, finding the following results:
$\mathcal{D}^{(N_f=2)}_{U(1)}(T)\vert_{inst} = 2\displaystyle\lim_{m\to 0}\chi_{top}(T)/m^2$ and
$\mathcal{D}^{(N_f=3)}_{U(1)}(T)\vert_{inst} = -6\displaystyle\lim_{m\to 0}\chi_{top}(T)/m^3$.

\subsubsection{Local $U(1)$ axial condensate}

Let us now consider the local condensate $\mathcal{C}_{U(1)}$ for $N_f=2$, which depends on the zero mode $\Psi_0(x,\rho,T)$ through the nontrivial integral $\int d^4x\,[\Psi_0\Psi_0^\dagger]^2$. For a finite-temperature instanton, the following expression holds for the zero mode (in the so-called \emph{singular gauge}) \cite{GPY1981,98InstQCD}:
\begin{equation}\label{0modetotalT}
\Psi^{\alpha,i}_0(x,\rho,T)=\psi_{0\mu}(x,\rho,T)v^{\alpha,i}_\mu,
\end{equation}
where the spinor $v^{\alpha,i}_\mu$ does not depend on $x$, $\rho$, $T$, and
\begin{equation}\label{0modeT}
\psi_{0\mu}(x,\rho,T) = \frac{1}{\pi\sqrt{2}\rho}\sqrt{\Pi}\,\partial_\mu\Big(\frac{\Phi}{\Pi}\Big),
\end{equation}
with
\begin{equation}\label{Pi-Phi}
\begin{split}
\Pi(x,\rho,T) &= 1+\frac{\pi\rho^2T}{r}\frac{\sinh{(2\pi T r)}}{\cosh{(2\pi T r)}-\cos{(2\pi T\tau)}},\\
\Phi(x,\rho,T) &= [\Pi(x,\rho,T)-1]\frac{\cos{(\pi T\tau)}}{\cosh{(\pi T r)}},
\end{split}
\end{equation}
$\tau$ being the Euclidean time and $r\equiv\sqrt{x^2_1+x^2_2+x^3_2}$.\\
Being interested only in the asymptotic temperature dependence of $\mathcal{C}_{U(1)}$, we shall neglect its structure with respect to the Dirac and colour indices, as it is irrelevant for our discussion. Let us now evaluate, making use of Eqs.~\eqref{0modetotalT}--\eqref{Pi-Phi}, the $T$ and $\rho$ dependence as $T\to\infty$ of the integral $\int d^4x\,[\psi_{0\mu}\psi_{0\mu}]^2$, which replaces the second Eq.~\eqref{normpsi0} in the integral \eqref{instcondensato2}. As $T\to\infty$, we obtain
\begin{equation*}
\Pi(x,\rho,T)\sim\frac{\pi\rho^2T}{r},\quad
\Phi(x,\rho,T)\sim\frac{\pi\rho^2T}{r}\cos{(\pi T\tau)}e^{-\pi T r},
\end{equation*}
and therefore, neglecting some constant multiplicative factors,
\begin{equation*}
\psi_{0\mu}(x,\rho,T) \sim \sqrt{\frac{T}{r}}\partial_\mu[\cos{(\pi T\tau)}e^{-\pi T r}]
\end{equation*}
(where we note that the dependence on $\rho$ has been cancelled), so that
\begin{equation}\label{0modelocal2}
\int d^4x\,[\psi_{0\mu}\psi_{0\mu}]^2\sim \int_{0}^{\beta}d\tau\int_{0}^{\infty} dr\, 4\pi r^2 \Big(\frac{T^3e^{-2\pi T r}}{r}\Big)^2 \sim 4\pi T^{5}\int_{0}^{\infty} dr\, e^{-4\pi T r}=T^4,
\end{equation}
while $\int d^4x\,\psi_{0\mu}\psi_{0\mu} \sim \textrm{const}$, as it must be.
Therefore, the expression \eqref{instcondensato2} for the local condensate $\mathcal{C}_{U(1)}$ in the instanton or anti-instanton background in the case $N_f=2$ is so modified at asymptotically high temperatures [using for $d_0(\rho,T)$ the expression \eqref{reduceddensityT}, which renders the integral convergent in the infrared, and for $g(\rho)$ the one-loop perturbative result \eqref{runningcoupling}]:
\begin{equation}\label{highTlocal}
\begin{split}
\mathcal{C}^{(N_f=2)}_{U(1)}(T)\vert_{inst}
&\sim T^4\int_{0}^{\infty} \frac{d\rho}{\rho^3}\, d_0(\rho,T) \sim T^4\int_{0}^{\infty} \frac{d\rho}{\rho^3}\, (\rho\Lambda_{QCD})^{b_0}e^{-\frac{2}{3}(\pi\rho T)^2(N_c+1)}\\
&=T^4(\pi T)^{2-b_0}\Lambda^{b_0}_{QCD}\int_{0}^{\infty} d\lambda\, \lambda^{b_0-3}e^{-\frac{2}{3}\lambda^2(N_c+1)},
\end{split}
\end{equation}
where the remaining integral over the dimensionless variable $\lambda$ yields a constant multiplicative factor. In other words, the asymptotic temperature dependence of $\mathcal{C}_{U(1)}$, for $N_f=2$ light flavours, is
\begin{equation}\label{localhighT}
\mathcal{C}^{(N_f=2)}_{U(1)}(T) \sim T^{6-b_0}.
\end{equation}
(Let us observe that, for $N_f=2$, $\mathcal{C}_{U(1)}$ has dimension $6$ in natural units.)\\
In conclusion, in the instanton-background approximation and in the case $N_f=2$, the local condensate $\mathcal{C}_{U(1)}$ is different from zero at any finite temperature $T$, and it vanishes asymptotically as $T\to\infty$, according to Eq. \eqref{localhighT}: that is, the $U(1)_A$ symmetry is effectively restored only asymptotically as $T\to\infty$.

Finally, let us point out the difficulties of extending this estimation to the case of an arbitrary number $N_f$ of light flavours. In this case, since $\mathcal{O}_{U(1)}$ consists of $2N_f$ quark fields, instead of Eq.~\eqref{0modelocal2} we find the following integral:
\begin{equation*}
\int d^4x\,[\psi_{0\mu}\psi_{0\mu}]^{N_f}\sim T^{3N_f-1}\int_{0}^{\infty} dr\, 4\pi r^{2-N_f} e^{-2N_f\pi T r},
\end{equation*}
which is clearly divergent in $r=0$ for $N_f>2$. While this approximate procedure gave a correct result for $N_f=2$, it cannot be used in the general case, making it necessary to perform a more precise and in-depth analysis. In fact, we also observe that a simple temperature dependence with a power law of the kind $\mathcal{C}_{U(1)}(T) \sim T^{3N_f-b_0} = T^{\frac{11}{3}(N_f-N_c)}$, obtained by naively extending the result \eqref{localhighT}, using purely dimensional considerations, would imply an increase (rather than a decrease) with the temperature in the case $N_f>N_c$: since the instantons are expected to be suppressed as $T\to\infty$, this divergent high-temperature behaviour of the $U(1)$ axial condensate appears to be really unnatural.\\
This problem could actually require some kind of nonperturbative renormalization: indeed, let us observe that, at $T=0$, being
\begin{equation*}
\int d^4x\,[\psi_0(x,\rho)]^{2N_f} = 2\pi^2\int_{0}^{\infty}dr\,r^3\Big[\frac{\sqrt{2}}{\pi}\frac{\rho}{(\rho^2+r^2)^{3/2}}\Big]^{2N_f} \propto \frac{1}{\rho^{4(N_f-1)}},
\end{equation*}
where $\psi_0(x,\rho)$ is given by Eq.~\eqref{0mode}, the result \eqref{instcondensato2} is generalized to the case of $N_f$ light flavours as follows:
\begin{equation*}
\mathcal{C}_{U(1)}\vert_{inst} \sim \int_{0}^{\infty}\frac{d\rho}{\rho^{3N_f+1}}\, d_0(\rho).
\end{equation*}
For $N_f\geq N_c$, this integral is affected by an ultraviolet divergence, instead of an infrared one: indeed, at small instanton sizes, the integrand diverges as $\rho^{b_0-(3N_f+1)}$. We observe that this UV divergence cannot be removed by the suppression factor in Eq.~\eqref{reduceddensityT} (which introduces only an infrared cutoff), and it resembles the one obtained in Refs. \cite{Shuryak1987,Shuryak1988}, where the density of instantonic molecules was found to be affected by a power UV divergence of nonperturbative nature. To overcome this problem, one could perhaps introduce some kind of nonperturbative renormalization (as suggested in the above-mentioned references) or some UV cutoff on the instanton size.

\section{Summary of the results and conclusions}

In this conclusive section we summarize our results and draw our conclusions, discussing also some possible future developments. The aim of our work was to study, in the chirally restored phase, some possible \emph{local} and \emph{global} genuine $U(1)$ axial condensates and their relations with the chiral susceptibilities \eqref{meson}--\eqref{chiralsusc} as well as with the topological susceptibility $\chi_{top} \equiv \langle Q^2 \rangle/V$, by means of two nonperturbative analytical techniques:\\
(i) by expressing the functional averages $\langle\ldots\rangle$ in terms of the spectral density $\rho(\lambda)$ of the Euclidean Dirac operator $i\slashed{D}$ and (ii) by evaluating the functional integrals in the instanton-background approximation.

In the Appendix, we have briefly reviewed (for the benefit of the reader) the most relevant results found in the past literature concerning the analysis of the chiral susceptibilities through the spectral density, and we have also extended some of the results to the case of an arbitrary number of light flavours $N_f$.
In particular, we have found that the relation $\chi_\pi-\chi_\delta\propto\displaystyle\lim_{m\to 0}\chi_{top}/m^2$ [see Eq. \eqref{chitopN}] still holds, above $T_c$ (in the chiral limit $m\to 0$), also for an arbitrary $N_f$:
by virtue of the well-known DIGA prediction for the quark-mass dependence $\chi_{top} = \mathcal{O}(m^{N_f})$ in the chiral limit $m\to 0$ at high temperatures $T$ \cite{GPY1981} (the same prediction is also derived using chiral effective Lagrangian models above $T_c$ \cite{EM2019,BM2020}), this result shows that, while for $N_f=2$ $\chi_\pi-\chi_\delta$ represents a good order parameter of the $U(1)_A$ symmetry, for $N_f>2$ the chiral susceptibilities become $U(1)_A$ symmetric at high temperatures ($\chi_\pi-\chi_\delta\to0$). Indeed, in the chirally restored phase ($T>T_c$), the $U(1)_A$ anomaly is expected to affect only the $2n$-point correlation functions with $n\geq N_f$ \cite{96Evans,96Lee,BCM1996}: the $U(1)$ axial condensates considered in this paper, being functional averages of $2N_f$-quark operators, are, therefore, the right objects to be studied in order to assess the $U(1)_A$ breaking.

The \textit{local} $U(1)$ axial condensates considered in this paper are functional averages $\mathcal{C}_{U(1)}\equiv\langle \mathcal{O}_{U(1)}(x)\rangle$ of \textit{local} $2N_f$-quark operators of the form \eqref{U1det} [see also Eqs. \eqref{local-operator_Nf=2} and \eqref{F-tensor} for the particular case $N_f=2$]: these operators are invariant under the whole chiral group except for the $U(1)_A$ transformations, so that, differently from the chiral condensate $\Sigma$ and the quantity $\chi_\pi-\chi_\delta$, their functional averages $\mathcal{C}_{U(1)}$ are ``genuine'' order parameters for the $U(1)_A$ symmetry alone (for any number of flavours) \cite{EM1994,Shuryak1994}.\\
In this paper, we have also proposed a new \textit{global} $U(1)$ axial condensate $\mathcal{D}_{U(1)}$, which is obtained taking the functional average of a \textit{multilocal} operator $\mathcal{O}_{U(1)}(x_1,\ldots,x_{N_f})$ of the form \eqref{multilocal-operator} [see also Eq. \eqref{multilocal-operator_Nf=2} for the particular case $N_f=2$] and then integrating over the four-space coordinates [see Eqs. \eqref{U1global_Nf=2} and \eqref{U1global}]:
the multilocal operator transforms under the chiral group exactly as the local operator $\mathcal{O}_{U(1)}(x)$, so that also $\mathcal{D}_{U(1)}$ is a genuine order parameter of the $U(1)_A$ symmetry. Moreover, its global feature renders it somewhat similar to the chiral susceptibilities.
The main motivation for introducing this new condensate is that [differently from the local $U(1)$ axial condensates] it can be studied using the spectral-density technique, as we have shown in detail in section 2.
[Moreover, the global feature of this new $U(1)$ axial condensate renders it a promising object for future numerical studies on the lattice, probably better than the local counterparts, for which, on the contrary, a direct numerical determination on the lattice is expected to be highly problematic; see, e.g., Ref. \cite{DM1995}.]

In particular, in section 2, we have derived, in the case $N_f=2$, the exact relation \eqref{U1global_Nf=2_final} between the global $U(1)$ axial condensate $\mathcal{D}_{U(1)}$ and the connected and disconnected components of the chiral susceptibilities, from which it follows that, in the chirally restored phase ($T>T_c$, $m=0$) where Eq.~\eqref{SU2rest} holds, $\mathcal{D}_{U(1)}$ turns out to be proportional to the $U(1)_A$-breaking difference $\chi_\pi-\chi_\delta$, and, thus, to $\displaystyle\lim_{m\to 0}\chi_{top}/m^2$ [see Eq. \eqref{result}].\\
However, while the chiral susceptibilities cease to reveal the $U(1)_A$ breaking if $N_f>2$, $\mathcal{D}_{U(1)}$ represents a good order parameter for any $N_f$: we have found that, for any number $N_f$ of light flavours, $\mathcal{D}_{U(1)}$ always contains a term proportional to $\chi_{top}/m^{N_f}$ [see Eq. \eqref{termN1}], which (by virtue of the above-mentioned DIGA prediction for the quark-mass dependence of $\chi_{top}$) is expected to remain different from zero in the chiral limit $m\to 0$ at high temperatures.

In section 3, we have explicitly computed the local and global $U(1)$ axial condensates in the high-temperature phase, using the instanton-background approximation of the functional integral. In this way, besides proving that these condensates are indeed different from zero in the high-temperature regime, we have also derived their asymptotic temperature dependence and compared it with that of the chiral and topological susceptibilities.
In particular, for an arbitrary number of light flavours $N_f$, we have found a nonzero result for the global condensate $\mathcal{D}_{U(1)}$ at high temperatures and in the chiral limit $m\to 0$, with the following asymptotic temperature dependence [see Eq. \eqref{U1global_T}]:
$\mathcal{D}_{U(1)}(T) \sim T^{4-N_f-b_0}$, where $b_0=(11N_c-2N_f)/3$.
Consistently with the fact that the instantons are asymptotically suppressed at high temperatures, the $U(1)$ axial condensate $\mathcal{D}_{U(1)}$ vanishes as $T\to\infty$, but it is anyhow different from zero at any finite temperature.\\
Instead, for the chiral susceptibilities, the following corresponding result has been found [see Eq. \eqref{suscT}]:
$(\chi_\pi-\chi_\delta)(T) \sim T^{2-b_0} (m/T)^{N_f-2}$,
which vanishes, in the case $N_f>2$, as $m^{N_f-2}$ in the chiral limit $m\to 0$, while it is nonzero in the particular case $N_f=2$.\\
In the case $N_f=2$, recalling that for $T>T_c$ (in the chiral limit $m\to 0$) $\mathcal{D}_{U(1)}$ comes out to be proportional to $\chi_{\pi}-\chi_{\delta}$ and to $\displaystyle\lim_{m\to 0}\chi_{top}/m^2$ [see Eq.~\eqref{result}], one also obtains that [see Eq. \eqref{chitopT_Nf=2}]
$\chi_{top}(T) \propto m^2\mathcal{D}_{U(1)}(T) \sim T^{4-b_0} (m/T)^2$,
which coincides with the well-known DIGA prediction for the asymptotic temperature dependence of $\chi_{top}$ \cite{GPY1981} in the case of $N_f=2$ light flavours (and which has been verified on the lattice: see, e.g., Refs. \cite{toplat1,toplat2,toplat3,toplat4,toplat5}; see also Ref. \cite{DDSX2017} for a recent review).\\
More in general, as we have already said, in section 2 we have shown that, for an arbitrary $N_f$, the $U(1)$ axial condensate $\mathcal{D}_{U(1)}$ contains a term proportional to $\chi_{top}/m^{N_f}$ [see Eq. \eqref{termN1}], and we have guessed that the whole global condensate $\mathcal{D}_{U(1)}$ might be proportional to $\displaystyle\lim_{m\to 0}\chi_{top}/m^{N_f}$ above the chiral phase transition, in the chiral limit $m\to 0$: indeed, the above-reported result for the asymptotic temperature dependence of $\mathcal{D}_{U(1)}$ is consistent with this guess and with the above-mentioned DIGA prediction for $\chi_{top}$ \cite{GPY1981}, i.e. [see Eq. \eqref{chitopT}],
$\chi_{top}(T) \propto m^{N_f}\mathcal{D}_{U(1)}(T) \sim T^{4-b_0} (m/T)^{N_f}$.

Finally, making use of the finite-temperature expression for the instanton zero mode, we have also estimated the asymptotic temperature dependence of the local $U(1)$ axial condensate in the case of $N_f=2$ light flavours, obtaining the result [see Eq. \eqref{localhighT}] $\mathcal{C}_{U(1)}\sim T^{6-b_0}$. (Instead, the case $N_f>2$ requires a more accurate and in-depth analysis and eventually the introduction of some kind of nonperturbative renormalization \cite{Shuryak1987,Shuryak1988} to remove a possible UV divergence that, as in the case at zero temperature, affects the result for more than two flavours: this issue, together with a more general and in-depth analysis of the renormalization of the \emph{local} and \emph{multilocal} operators discussed in this paper, is left to future works.)

In conclusion, making use of the instanton-background approximation, we have found that these $2N_f$-quark $U(1)$ axial condensates are different from zero at any finite temperature $T$ and vanish only asymptotically as $T\to\infty$.\\
As we have seen, for $N_f>2$, the $U(1)_A$ anomaly seems to have no effects on the chiral susceptibilities above $T_c$, so that an effective restoration of the $U(1)_A$ symmetry at the level of the $q\overline{q}$ scalar and pseudoscalar meson mass spectrum is expected, in spite of the fact that the $U(1)_A$ symmetry is manifestly broken by the above-mentioned $2N_f$-quark condensates (see also Ref. \cite{MM2013}).\\
Instead, for the (physically most relevant) case $N_f=2$, the global $U(1)$ axial condensate $\mathcal{D}_{U(1)}$ comes out to be proportional to $\chi_\pi - \chi_\delta$ for $T>T_c$, so that, $\mathcal{D}_{U(1)}$ being nonzero above $T_c$, an effective restoration of the $U(1)_A$ symmetry at the level of the $q\overline{q}$ scalar and pseudoscalar meson mass spectrum above $T_c$ is excluded.\\
Of course, we cannot make more quantitative statements about the real magnitude of these condensates and so of the breaking of the $U(1)_A$ symmetry (which could be obtained only by numerical calculations), so that we cannot exclude that an \emph{approximate} restoration of the $U(1)_A$ symmetry may anyhow happen in the vicinity of $T_c$.
Numerical studies of these $U(1)$ axial condensates on the lattice (and also a direct investigation of the quark-mass dependence of $\chi_{top}$ above the chiral transition) could, of course, allow for a deeper understanding of this important problem and (hopefully) provide a first-principle confirmation of the analytical results found in this paper: the fact that, as we have already said, present lattice data are in agreement with the DIGA prediction for the asymptotic temperature dependence of $\chi_{top}$ already gives, however, indirect support to the results found in this paper concerning the $U(1)$ axial condensates.


\newpage

\renewcommand{\thesection}{}
\renewcommand{\thesubsection}{A.\arabic{subsection} }

\pagebreak[3]
\setcounter{section}{1}
\setcounter{equation}{0}
\setcounter{subsection}{0}
\setcounter{footnote}{0}

\begin{flushleft}
{\Large\bf \thesection Appendix: A review on chiral susceptibilities and their spectral-density analysis}
\end{flushleft}

\renewcommand{\thesection}{A}

\subsection{Chiral susceptibilities in QCD with $N_f=2$ light flavours}

In this appendix we shall review, for the benefit of the reader, the most relevant relations for the chiral susceptibilities of all the meson channels in terms of the so-called \emph{spectral density} (see, e.g., Refs. \cite{SmilgaBook,82Floratos,93Smilga}) of the Euclidean Dirac operator $i\slashed{D}[A] = i\gamma_{E\mu} D_\mu$, where $D_\mu = \partial_\mu + i g A_\mu$ is the covariant derivative in a given gauge field $A_\mu$ and $\gamma_{E\mu}$ ($\mu = 0,1,2,3$) are the Hermitian Euclidean gamma matrices:\footnote{We use the Hermitian Euclidean gamma matrices $\gamma_{E\mu}$, related to the Minkowskian gamma matrices $\gamma^\mu$ by the relations $\gamma_{E0} = \gamma^0$ and $\gamma_{Ek} = -i\gamma^k$ ($k=1,2,3$), so that they satisfy the condition: $\{ \gamma_{E\mu} \gamma_{E\nu} \} = 2\delta_{\mu\nu} {\bf I}_{4 \times 4}$. We also use the following definition: $\gamma_5 \equiv -\gamma_{E0}\gamma_{E1}\gamma_{E2}\gamma_{E3} = -i\gamma^0\gamma^1\gamma^2\gamma^3$.}
\begin{equation}\label{density}
\rho(\lambda) \equiv \langle \frac{1}{V}\sum_{k} \delta(\lambda-\lambda_k[A]) \rangle .
\end{equation}
$\{\lambda_k[A]\}$ is the set of (real) gauge-field-dependent eigenvalues of $i\slashed{D}[A]$, i.e., $i\slashed{D} u_k = \lambda_k u_k$, $u_k$ being the corresponding eigenvectors, and the brackets at the right-hand side stand for the following functional integration over the gauge field:
\begin{equation}\label{funcaverage}
\langle\dots\rangle=\frac{1}{Z}\int DA\,e^{-S_G[A]}\,[\det(\slashed{D}[A]+m)]^{N_f}\dots,
\end{equation}
where $S_G[A]$ is the pure-gauge (Yang-Mills) action and $m$ is the quark mass, which we assume, for simplicity, to be equal for all the $N_f$ quark flavours (the chiral limit $m\to 0$ will be taken later on).

The key ingredient which allows us to relate the chiral susceptibilities with the spectral density $\rho(\lambda)$ is the so-called \emph{spectral decomposition} of the quark propagator $\mathcal{G}_A(x,y)$ in an external gauge field $A_\mu$ \cite{SmilgaBook,82Floratos,93Smilga}:
\begin{equation}\label{prop}
\mathcal{G}^{\alpha\beta,ij}_{A,ff'}(x,y) \equiv \langle q^{\alpha,i}_f(x)\overline{q}^{\beta,j}_{f'}(y)\rangle_A = \delta_{ff'}\sum_k \frac{u^{\alpha,i}_k(x) u^{\dagger\beta,j}_k(y)}{m-i\lambda_k},
\end{equation}
where $f$ and $f'$ are flavour indices (note that, since we are assuming equal quark masses, $\mathcal{G}_A$ is proportional to the identity matrix in flavour space), $\alpha$ and $\beta$ are colour indices, $i$ and $j$ are Dirac indices, and $\langle \dots \rangle_A$ stands for the functional integration over the quark fields only, in a given gauge field $A_\mu$.

Let us start with the chiral susceptibility in the isoscalar ($I=0$) scalar channel $J_{\sigma}=\overline{q}q$ (the relations that we shall derive can be also found, for example, in Refs. \cite{lat2012,lat2014,12Ohno}):
\begin{equation}\label{chisigmadef}
\chi_{\sigma} = \frac{1}{V}\int d^4x\int d^4y\, \langle J_{\sigma}(x)J^{\dagger}_{\sigma}(y)\rangle.
\end{equation}
Explicitly performing the functional integration over the quark fields, one obtains
\begin{equation}\label{contisigma}
\begin{split}
\langle J_\sigma(x)J^{\dagger}_\sigma(y)\rangle =
-\langle \Tr[\mathcal{G}_A(x,y)\mathcal{G}_A(y,x)]\rangle + \langle \Tr\mathcal{G}_A(x,x)\Tr\mathcal{G}_A(y,y)\rangle ,
\end{split}
\end{equation}
where the traces are done with respect to Dirac, colour, and flavour indices, and the brackets at the right-hand side stand for the functional integration \eqref{funcaverage}.\\
Substituting this expression into Eq. \eqref{chisigmadef}, one obtains
\begin{equation}\label{chisigma}
\chi_\sigma = \chi_{\sigma,conn} + \chi_{\sigma,disc},
\end{equation}
where $\chi_{\sigma,conn}$ and $\chi_{\sigma,disc}$ are defined as the \textit{connected} and \textit{disconnected} components, respectively, of $\chi_\sigma$, with respect to the quark lines in an external field (that is, \textit{before} averaging over the gauge fields):
\begin{equation}\label{sigma_conn,disc}
\begin{split}
\chi_{\sigma,conn} &\equiv -\frac{1}{V}\int d^4x\int d^4y\,\langle \Tr[\mathcal{G}_A(x,y)\mathcal{G}_A(y,x)]\rangle,\\
\chi_{\sigma,disc} &\equiv \frac{1}{V}\int d^4x\int d^4y\,\langle \Tr\mathcal{G}_A(x,x)\Tr\mathcal{G}_A(y,y)\rangle.
\end{split}
\end{equation}
By using the orthonormality condition for the eigenfunctions $u_k$,
\begin{equation}\label{orthonorm}
\int d^4x\, u^{\dagger\alpha,i}_k(x) u^{\alpha,i}_{k'}(x) = \delta_{kk'},
\end{equation}
and the well-known fact that $\rho(\lambda)$ is an even function of $\lambda$ [$\rho(-\lambda) = \rho(\lambda)$], one can easily express the connected term in terms of the spectral density:
\begin{equation}\label{sigmaconn}
\chi_{\sigma,conn} = -\frac{2}{V}\langle \sum_k\frac{1}{(m-i\lambda_k)^2}\rangle = -4\int_{0}^{\infty}d\lambda\,\rho(\lambda)\frac{m^2-\lambda^2}{(m^2+\lambda^2)^2}.
\end{equation}
Instead, it is not possible to repeat the same calculations for the disconnected part. In fact, by using the orthonormality condition \eqref{orthonorm}, one obtains two independent summations which cannot be directly written in terms of $\rho(\lambda)$:
\begin{equation}\label{sigmadiscaltro}
\chi_{\sigma,disc} = \frac{4}{V}\langle \sum_k\frac{1}{m-i\lambda_k}\sum_{k'}\frac{1}{m-i\lambda_{k'}}\rangle.
\end{equation}
Nevertheless, it is possible to relate $\chi_{\sigma,disc}$ to a derivative of the spectral density. In fact, the total susceptibility $\chi_\sigma$ is related to the chiral condensate $\Sigma = -\langle \overline{q}q \rangle$ as follows:
\begin{equation}\label{sigmaSigmaTrue}
\chi_\sigma = \frac{\partial \Sigma}{\partial m}+V\Sigma^2,
\end{equation}
being $\Sigma=\frac{1}{V}\frac{\partial\ln Z}{\partial m}$. Since $\Sigma=-\langle J_\sigma(0)\rangle $, Eq.~\eqref{sigmaSigmaTrue} can be rewritten as
\begin{equation*}
\int d^4x\,[\langle J_\sigma(x)J_\sigma(0)\rangle-\langle J_\sigma(x)\rangle\langle J_\sigma(0)\rangle]=\frac{\partial \Sigma}{\partial m}.
\end{equation*}
It is convenient to redefine the chiral susceptibility as the left-hand side of this expression, i.e., by subtracting the disconnected product of the two expectation values $\langle J_\sigma(x)\rangle\langle J_\sigma(0)\rangle$:\footnote{The chiral susceptibility redefined in this way is usually referred to as ``connected'' with respect to the full theory. Here, instead, we shall use the term ``connected'' as in Eq.~\eqref{sigma_conn,disc}, i.e., with respect to the quark lines in an external gauge field $A_\mu$.}
\begin{equation}\label{newsigma1}
\overline{\chi}_\sigma\equiv \int d^4x\,[\langle J_\sigma(x)J_\sigma(0)\rangle-\langle J_\sigma(x)\rangle\langle J_\sigma(0)\rangle],
\end{equation}
that is
\begin{equation}\label{newsigma}
\overline{\chi}_\sigma \equiv \chi_\sigma-V\Sigma^2 = \frac{\partial \Sigma}{\partial m}.
\end{equation}
We also define the connected and disconnected components of $\overline{\chi}_\sigma$ as, respectively,
\begin{equation*}
\begin{split}
\overline{\chi}_{\sigma,conn} &\equiv \chi_{\sigma,conn},\\
\overline{\chi}_{\sigma,disc} &\equiv \chi_{\sigma,disc}-V\Sigma^2.
\end{split}
\end{equation*}	
Clearly, above $T_c$ and in the chiral limit $m\to 0$, the chiral condensate $\Sigma$ vanishes and, therefore, $\chi_\sigma = \overline{\chi}_\sigma$. For convenience, since we shall mainly consider the chirally restored phase for $T>T_c$, in the following, we shall continue to use the notation $\chi_\sigma$. Making use of a well-known expression for the chiral condensate in terms of the spectral density \cite{SmilgaBook,93Smilga},
\begin{equation}\label{rhoSigma}
\Sigma = 2N_f m\int_{0}^{\infty} d\lambda\,\frac{\rho(\lambda)}{m^2+\lambda^2}
\end{equation}
[which in the chiral limit $m\to 0$ leads to the famous \emph{Banks-Casher relation} $\Sigma = N_f\pi \rho(0)$ \cite{80Banks}], we can rewrite Eq.~\eqref{newsigma} as follows:
\begin{equation*}
\chi_\sigma = \frac{\partial \Sigma}{\partial m} = 4\int_{0}^{\infty}d\lambda\,\rho(\lambda,m)\frac{\lambda^2-m^2}{(m^2+\lambda^2)^2} + 4\int_{0}^{\infty}d\lambda\,\frac{\partial\rho(\lambda,m)}{\partial m}\frac{m}{m^2+\lambda^2},
\end{equation*}
where the dependence of $\rho$ on the quark mass $m$ (coming from the weight $[\det(\slashed{D}+m)]^2$ in the functional integral over the gauge fields) has been explicitly indicated.
The first term is just equal to the connected susceptibility (\ref{sigmaconn}), and, therefore, it follows that
\begin{equation}\label{sigmadisc}
\chi_{\sigma,disc} = 4\int_{0}^{\infty}d\lambda\,\frac{\partial\rho(\lambda,m)}{\partial m}\frac{m}{m^2+\lambda^2}.
\end{equation}
Let us now consider the chiral susceptibility in the isovector ($I=1$) scalar channel $J^a_{\delta}=\overline{q}\tau_aq$, corresponding to the three mesons $\delta_a$:
\begin{equation*}
\chi_{\delta_a} = \frac{1}{V}\int d^4x\int d^4y\, \langle J^a_{\delta}(x)(J^a_{\delta}(y))^{\dagger}\rangle,
\end{equation*}
where, in this case,
\begin{equation*}
\begin{split}
\langle J^a_\delta(x)(J^b_\delta(y))^{\dagger}\rangle &= -\langle \Tr[\tau_a\mathcal{G}_A(x,y)\tau_b\mathcal{G}_A(y,x)]\rangle + \langle \Tr[\tau_a\mathcal{G}_A(x,x)]\Tr[\tau_b\mathcal{G}_A(y,y)]\rangle \\
&= -2\delta_{ab}\langle \Tr_{DC}[\mathcal{G}_A(x,y)\mathcal{G}_A(y,x)]\rangle.
\end{split}
\end{equation*}
In the last passage, the trace over the flavour indices has been explicitly performed: let us observe that in this case the disconnected term vanishes, because $\Tr(\tau_a)=0$, and, using $\Tr(\tau_a\tau_b)=2\delta_{ab}$, we are left with the above-reported expression for the connected term, having indicated with ``$\Tr_{DC}$'' the trace over the Dirac and colour indices only and with $\mathcal{G}_A$ the quark propagator \eqref{prop} with $f=f'$.
Of course, the isospin symmetry (the quark masses being equal to the common value $m$) implies that $\chi_{\delta_a}$ is the same for $a=1,2,3$, and moreover, from a comparison with Eq.~\eqref{contisigma}, it is easy to see that $\chi_{\delta}\equiv\chi_{\delta_a}=\chi_{\sigma,conn}$.

The isoscalar ($I=0$) pseudoscalar channel $J_\eta=i\overline{q}\gamma_5q$ is associated to the singlet $\eta$, which is the two-flavour version of the meson $\eta'$. As in the scalar channel, the integration over the quark fields gives rise to a connected and a disconnected term:
\begin{equation}\label{contieta}
\langle J_\eta(x)J^{\dagger}_\eta(y)\rangle = \langle \Tr[\gamma_5\mathcal{G}_A(x,y)\gamma_5\mathcal{G}_A(y,x)]\rangle - \langle \Tr[\gamma_5\mathcal{G}_A(x,x)]\Tr[\gamma_5\mathcal{G}_A(y,y)]\rangle .
\end{equation}
The chiral susceptibility $\chi_\eta$ is thus given by
$\chi_\eta = \chi_{\eta,conn}+\chi_{\eta,disc}$,
where the connected and disconnected terms (with respect to the quark lines in an external gauge field) are defined analogously to Eq.~\eqref{sigma_conn,disc}, i.e.,
\begin{equation}\label{eta_conn,disc}
\begin{split}
\chi_{\eta,conn} &\equiv \frac{1}{V}\int d^4x\int d^4y\,\langle \Tr[\gamma_5\mathcal{G}_A(x,y)\gamma_5\mathcal{G}_A(y,x)]\rangle,\\
\chi_{\eta,disc} &\equiv -\frac{1}{V}\int d^4x\int d^4y\,\langle \Tr[\gamma_5\mathcal{G}_A(x,x)]\Tr[\gamma_5\mathcal{G}_A(y,y)]\rangle.
\end{split}
\end{equation}
By using the fact that $\gamma_5\mathcal{G}_A(x,y)\gamma_5=\mathcal{G}^{\dagger}_A(y,x)$ \cite{SmilgaBook}, the connected part can be straightforwardly written in terms of the spectral density:
\begin{equation}\label{etaconn}
\chi_{\eta,conn} = \frac{2}{V}\langle \sum_k\frac{1}{m^2+\lambda_k^2}\rangle = 4\int_{0}^{\infty}d\lambda\,\frac{\rho(\lambda)}{m^2+\lambda^2} = \frac{\Sigma}{m}.
\end{equation}
Instead, it is not possible to express the disconnected part straightforwardly in terms of the spectral density. Nevertheless, one can derive an important relation between $\chi_{\eta,disc}$ and the topological susceptibility $\chi_{top}$, which is defined in Euclidean space as
\begin{equation}\label{chitop}
\chi_{top} \equiv \int d^4x\,\langle q(x)q(0)\rangle = \frac{1}{V}\int d^4x\int d^4y\,\langle q(x)q(y)\rangle=\frac{1}{V}\langle Q^2\rangle,
\end{equation}
where $q(x) = \frac{g^2}{64\pi^2}\varepsilon_{\mu\nu\rho\sigma}F^a_{\mu\nu}F^a_{\rho\sigma}$ (with $\varepsilon_{0123}=-1$) is the topological charge density and $Q=\int d^4x\, q(x)$ is the topological charge. As a consequence of the fact that the Euclidean Dirac operator $i\slashed{D}$ anticommutes with the $\gamma_5$ matrix, it is easy to see that $\chi_{\eta,disc}$ gets a contribution only from the so-called \emph{zero modes} of the Euclidean Dirac operator,
\begin{eqnarray}\label{etadisc}
\chi_{\eta,disc} &=& -\frac{4}{V}\langle \int d^4x\int d^4y\,\Tr_{DC}\left[ \sum_k \frac{\gamma_5 u_k(x) u^\dagger_k(x)}{m-i\lambda_k} \right] \Tr_{DC}\left[ \sum_{k'} \frac{\gamma_5 u_{k'}(y) u^\dagger_{k'}(y)}{m-i\lambda_{k'}} \right] \rangle \nonumber\\
&=& -\frac{4}{V}\langle\frac{1}{m^2}(n_L-n_R)^2\rangle = -\frac{4}{m^2}\frac{1}{V}\langle Q^2\rangle = -4\frac{\chi_{top}}{m^2},
\end{eqnarray}
where $n_L$ and $n_R$ are the number of \emph{left-handed} and \emph{right-handed} zero modes respectively (i.e., $\gamma_5 u_{L,R} = \pm u_{L,R}$), which, as is well known, are related to the topological charge $Q$ of the given gauge-field configuration by the \emph{Atiyah-Singer theorem} $Q=n_R-n_L$ \cite{84Atiyah}.

Finally, let us consider the isovector ($I=1$) pseudoscalar channel $J^a_\pi=i\overline{q}\gamma_5\tau_aq$, corresponding to the three pions $\pi_a$. In the chiral limit $m\to 0$ and in the low-temperature phase ($T<T_c$), these are the Goldstone bosons generated by the chiral symmetry breaking of $SU(2)_V \otimes SU(2)_A$ down to $SU(2)_V$. As for the isovector ($I=1$) scalar channel, the chiral susceptibility $\chi_\pi$ consists of the connected term only and, moreover, it is equal to the connected part of $\chi_\eta$:
\begin{equation}\label{pi}
\chi_{\pi} \equiv \chi_{\pi_a} = \chi_{\eta,conn} =\frac{\Sigma}{m},
\end{equation}
as is evident from a comparison of the expression
\begin{equation}
\langle J^a_\pi(x)(J^b_\pi(y))^{\dagger}\rangle = \langle \Tr[\gamma_5\tau_a\mathcal{G}_A(x,y)\gamma_5\tau_b\mathcal{G}_A(y,x)]\rangle = 2\delta_{ab}\langle \Tr_{DC}[\gamma_5\mathcal{G}_A(x,y)\gamma_5\mathcal{G}_A(y,x)]\rangle
\end{equation}
with Eqs.~\eqref{contieta}--\eqref{etaconn}. Let us observe that in the chiral limit $m\to 0$ and for $T<T_c$ the chiral condensate $\Sigma$ is nonzero and, therefore, $\chi_\pi$ diverges, which is consistent with the fact that the pions are massless in this case.\footnote{Indeed, $\langle J_M(x)J^{\dagger}_M(0)\rangle-\langle J_M(x)\rangle\langle J^{\dagger}_M(0)\rangle\sim e^{-\mathcal{M}_M|x|}$ as $x\to\infty$, where $\mathcal{M}_M$ is the \emph{screening mass} of the meson excitation $M$ interpolated by the operator $J_M$ \cite{DK1987}. We note that $\langle J_M\rangle$ vanishes in all the meson channels except for the $\sigma$ one, in which case it vanishes only in the chirally restored phase $T>T_c$.}

Let us summarize the expression obtained so far:
\begin{align}
\chi_\sigma&=\chi_{\sigma,conn}+\chi_{\sigma,disc}, & \chi_{\sigma,conn}&=-4\int_{0}^{\infty}d\lambda\,\rho(\lambda,m)\frac{m^2-\lambda^2}{(m^2+\lambda^2)^2},\nonumber\\
\chi_\delta&=\chi_{\sigma,conn},	& \chi_{\sigma,disc}&=4\int_{0}^{\infty}d\lambda\,\frac{\partial\rho(\lambda,m)}{\partial m}\frac{m}{m^2+\lambda^2},\nonumber\\
\chi_\eta&=\chi_{\eta,conn}+\chi_{\eta,disc}, & \chi_{\eta,conn}&=\frac{\Sigma}{m}=4\int_{0}^{\infty}d\lambda\,\frac{\rho(\lambda,m)}{m^2+\lambda^2},\nonumber\\
\chi_\pi&=\chi_{\eta,conn}, & \chi_{\eta,disc}&=-4\frac{\chi_{top}}{m^2}.\label{chiralsusc2}
\end{align}
In the chiral limit $m\to 0$ and for $T>T_c$, the $SU(2)_V \otimes SU(2)_A$ chiral restoration implies that [see Eq. \eqref{symmpict}]
\begin{equation}\label{SU2rest1}
\begin{split}
\chi_\pi-\chi_\sigma&=\chi_\pi-\chi_\delta-\chi_{\sigma,disc}=0,\\
\chi_\eta-\chi_\delta&=\chi_\pi-\chi_\delta-4\lim_{m\to 0}\frac{\chi_{top}}{m^2}=0.
\end{split}
\end{equation}
Therefore, in the chirally restored phase, we can relate the $U(1)_A$-breaking difference $\chi_\pi-\chi_\delta=\chi_{\eta,conn}-\chi_{\sigma,conn}$ to the topological susceptibility $\chi_{top}$:
\begin{equation}\label{SU2rest}
\chi_\pi-\chi_\delta=\chi_{\eta,conn}-\chi_{\sigma,conn}=\chi_{\sigma,disc}=-\chi_{\eta,disc}=4\lim_{m\to 0}\frac{\chi_{top}}{m^2}.
\end{equation}
We stress that this equation must hold for every $T>T_c$, as long as we take the chiral limit $m\to 0$. If the whole chiral group $U(2)_V \otimes U(2)_A$ were realized \emph{\`a la} Wigner-Weyl, all the chiral susceptibility should be equal and each member of Eq.~\eqref{SU2rest} should vanish in the chiral limit.
However, in the so-called \emph{dilute instanton gas approximation} (DIGA) \cite{GPY1981}, the topological susceptibility $\chi_{top}$ at high temperatures is expected to be of the order of $\mathcal{O}(m^{N_f})$, in the chiral limit $m\to 0$: as a consequence, being here $N_f=2$, we expect that the right-hand side of Eq. \eqref{SU2rest} is nonzero and, therefore, that the $U(1)_A$ symmetry remains manifestly broken even at high $T$ and that the whole chiral group $U(2)_V \otimes U(2)_A$ is effectively restored only asymptotically as $T\to\infty$.

Finally, let us briefly discuss the expression of the $U(1)_A$-breaking difference $\chi_\pi-\chi_\delta$ in terms of the spectral density $\rho$. From Eq.~\eqref{chiralsusc2}, one finds that
\begin{equation} \label{U1order_Nf=2}
\chi_\pi-\chi_\delta=8\int_{0}^{\infty}d\lambda\,\frac{m^2\rho(\lambda,m)}{(m^2+\lambda^2)^2}=\frac{8}{m}\int_0^\infty dz\,\frac{\rho(mz,m)}{(1+z^2)^2}.
\end{equation}
Analogously to the case of the $SU(2)_V \otimes SU(2)_A$ symmetry, also an eventual $U(1)_A$ breaking is expected to come from the low-lying part of the Dirac spectrum. However, in order for $\chi_\pi-\chi_{\delta}$ to vanish in the chiral limit $m\to 0$, it is not sufficient that $\rho(\lambda=0,m=0)=0$, as for the chiral condensate $\Sigma$, but also the behaviour of $\rho(\lambda,m)$ as $\lambda\to 0$ and $m\to0$ is important. For example, as noted in Refs. \cite{lat2012,lat2014}, if $\rho(\lambda,m)\sim m^{\nu_m}|\lambda|^{\nu_{\lambda}}$ with $\nu_m+\nu_\lambda=1$ as $m,\,\lambda\to0$, then, according to Eq. \eqref{U1order_Nf=2}, the $U(1)_A$ order parameter $\chi_{\pi}-\chi_{\delta}$ is nonzero in the chiral limit. In other words, proper forms of the function $\rho(\lambda,m)$ can result in scenarios in which the $U(1)_A$ symmetry is manifestly broken even though the $SU(2)_V \otimes SU(2)_A$ symmetry is restored. Another $U(1)_A$-broken scenario, which is particularly important to our discussion, is the one obtained by considering the contribution of the discrete Dirac zero modes, as we will discuss in the last subsection of this appendix.

\subsection{Chiral susceptibilities in QCD with $N_f$ light flavours}

In this subsection, we extend the study of the chiral susceptibilities to the case of an arbitrary number $N_f$ of light quark flavours (with a common mass $m$). In particular, we determine the new constraints imposed by the chiral restoration: one cannot straightforwardly extend the conditions \eqref{SU2rest1} and \eqref{SU2rest} of $SU(2)_V \otimes SU(2)_A$ restoration to the general case of $N_f$ flavours, because the mixing between the meson channels is now much more complicate than the one shown in Eq.~\eqref{symmpict}. The definitions \eqref{meson} of quark bilinears $J_M$ can be straightforwardly extended to an arbitrary number of flavour $N_f$: in the case $N_f=2$, $\tau_a$ (with $a=1,2,3$) are the Pauli matrices and the $\pi_a$ channel corresponds to the three pions, while in the more general case $\tau_a$ ($a=1,\ldots,N_f^2-1$) are the generators of $SU(N_f)$ in the fundamental representation, normalized as $\Tr(\tau_a\tau_b)=2\delta_{ab}$, and the $\pi_a$ form a multiplet of $N_f^2-1$ pseudoscalar mesons.

It is easy to see that the $U(1)_A$ transformations mix the meson channels in the same way as in Eq.~\eqref{symmpict}. Therefore, $\chi_{\pi}-\chi_{\delta}\neq0$ still implies $U(1)_A$ breaking (but the reverse implication may not be true). As regards the $SU(N_f)_A$ transformations, one can easily derive that
\begin{equation}\label{sigma-pi,eta-delta}
SU(N_f)_A:\quad J_\sigma =\overline{q}q\rightarrow J'_\sigma=J^3_\pi+J^{(N_f-2)}_\sigma,\quad J_\eta = i\overline{q}\gamma_5q\rightarrow J'_\eta=J^3_\delta+J^{(N_f-2)}_\eta,
\end{equation}
under a rotation of an angle $\theta=\pm\pi/2$, respectively, along the generator $\tau_3$, which is the third Pauli matrix in the subspace of the up and down quarks and zero elsewhere. We have defined $J^{(N_f-2)}_\sigma\equiv\overline{q}^{(N_f-2)}q^{(N_f-2)}=\overline{s}s+\ldots$, where $q^{(N_f-2)}$ has the first two components (the up and down) set equal to zero: thus, $J^{(N_f-2)}_\sigma$ has just $N_f-2$ terms. The bilinear $J^{(N_f-2)}_\eta$ is defined analogously. The restoration of $SU(N_f)_V \otimes SU(N_f)_A$ implies that
\begin{equation}\label{JJ}
\langle J_\sigma J_\sigma \rangle=\langle J'_\sigma J'_\sigma \rangle=\langle J^3_\pi J^3_\pi \rangle + \langle J^{(N_f-2)}_\sigma J^{(N_f-2)}_\sigma \rangle,
\end{equation}
where the mixed terms vanish, as they contain only one $SU(N_f)$ generator $\tau_3$. Once integrated over the quark fields, the correlation function $\langle J_\sigma J_\sigma \rangle$ consists of a connected and a disconnected term, provided with one or two traces over the quark indices, respectively: therefore, the first term gets a factor $N_f$ from the trace over the flavour indices, while the second one gets a factor $N_f^2$. Conversely, $\langle J^3_\pi J^3_\pi \rangle$ does not depend (explicitly) on $N_f$, since the trace over the flavour indices yields $\Tr(\tau_3^2)=2$. Let us rewrite the quantities involved in Eq.~\eqref{JJ} so that the dependence on $N_f$, coming from the trace operation, is made explicit:
\begin{equation}
\begin{split}
\langle J_\sigma J_\sigma \rangle&=N_f\langle \mathcal{A}\rangle+N_f^2\langle \mathcal{B}\rangle,\\
\langle J^{(N_f-2)}_\sigma J^{(N_f-2)}_\sigma \rangle&=(N_f-2)\langle \mathcal{A}\rangle+(N_f-2)^2\langle \mathcal{B}\rangle,\\
\langle J^3_\pi J^3_\pi \rangle&=\langle \mathcal{C}\rangle,
\end{split}
\end{equation}
where $\mathcal{A}, \mathcal{B}$, and $\mathcal{C}$ do not depend on $N_f$. (However, their functional average depends on $N_f$ through the determinant of the quark matrix $[\det(\slashed{D}+m)]^{N_f}$.) Substituting these expressions in Eq.~\eqref{JJ}, we obtain the following equation:
\begin{equation}\label{conditionN}
2\langle \mathcal{A}\rangle+4(N_f-1)\langle \mathcal{B}\rangle=\langle \mathcal{C}\rangle,
\end{equation}
which has to be satisfied if $SU(N_f)_V \otimes SU(N_f)_A$ is restored. In the particular case $N_f=2$, Eq.~\eqref{conditionN} becomes $2\langle \mathcal{A}\rangle+4\langle \mathcal{B}\rangle=\langle \mathcal{C}\rangle$, which is nothing else than $\langle J_\sigma J_\sigma \rangle=\langle J^3_\pi J^3_\pi \rangle$, and results in the constraint $\chi_\sigma=\chi_\pi$.

Before analysing the condition \eqref{conditionN} further, let us redefine the quark correlators in a more convenient manner. Indeed, we note that $\frac{1}{V}\int d^4x\int d^4y\,\langle \mathcal{A}\rangle=\frac{\chi_{\sigma,conn}}{N_f}$ and $\frac{1}{V}\int d^4x\int d^4y\,\langle \mathcal{B}\rangle=\frac{\chi_{\sigma,disc}}{N_f^2}$, $\chi_{\sigma,conn}$ and $\chi_{\sigma,disc}$ being the connected and disconnected susceptibilities, respectively, defined as in Eqs.~\eqref{chisigmadef}--\eqref{sigma_conn,disc}.

In order to eliminate these $N_f$ factors, it is convenient to renormalize the isoscalar channels by a factor $\sqrt{\frac{2}{N_f}}$, so that they are normalized as the isovector ones:
\begin{equation}\label{mesonN}
\tilde{J}_\sigma \equiv \overline{q}\tau_0q,\quad \tilde{J}_\eta \equiv i\overline{q}\gamma_5\tau_0q,
\end{equation}
where $\tau_0\equiv\sqrt{\frac{2}{N_f}}{\bf I}_{N_f \times N_f}$ and, thus, $\Tr(\tau_0^2)=\Tr(\tau_a^2)=2$. By redefining also the chiral susceptibilities $\tilde\chi_\sigma$ and $\tilde\chi_\eta$ with these renormalized isoscalar operators, we easily derive the following generalization of Eq.~\eqref{chiralsusc2}:
\begin{align}\label{chiralsuscN}
\tilde\chi_\sigma&=\tilde\chi_{\sigma,conn}+\tilde\chi_{\sigma,disc}, & \tilde\chi_{\sigma,conn}&=-4\int_{0}^{\infty}d\lambda\,\rho(\lambda,m)\frac{m^2-\lambda^2}{(m^2+\lambda^2)^2},\nonumber\\
\chi_\delta&=\tilde\chi_{\sigma,conn},	& \tilde\chi_{\sigma,disc}&=4\int_{0}^{\infty}d\lambda\,\frac{\partial\rho(\lambda,m)}{\partial m}\frac{m}{m^2+\lambda^2},\nonumber\\
\tilde\chi_\eta&=\tilde\chi_{\eta,conn}+\tilde\chi_{\eta,disc}, & \tilde\chi_{\eta,conn}&=\frac{2}{N_f}\frac{\Sigma}{m}=4\int_{0}^{\infty}d\lambda\,\frac{\rho(\lambda,m)}{m^2+\lambda^2},\nonumber\\
\chi_\pi&=\tilde\chi_{\eta,conn}, & \tilde\chi_{\eta,disc}&=-2N_f\frac{\chi_{top}}{m^2},
\end{align}
where $\Sigma$ is still given by Eq.~\eqref{rhoSigma}. Now, being $\tilde\chi_\sigma=\frac{2}{N_f}\chi_\sigma$, by integrating the condition \eqref{conditionN} over $\frac{1}{V}\int d^4x\int d^4y$, we obtain the following relation:
\begin{equation}\label{conditionN1}
\tilde\chi_{\sigma,conn}+\frac{2(N_f-1)}{N_f}\tilde\chi_{\sigma,disc}=\chi_\pi.
\end{equation}
Again, for $N_f=2$ this is just the degeneracy condition $\chi_\sigma=\chi_\pi$. Starting from Eq.~\eqref{sigma-pi,eta-delta}, we can repeat all the above passages also for the $\eta$ channel, obtaining
\begin{equation}\label{conditionN2}
\tilde\chi_{\eta,conn}+\frac{2(N_f-1)}{N_f}\tilde\chi_{\eta,disc}=\chi_\delta.
\end{equation}
We recall that this equation, together with Eq.~\eqref{conditionN1}, must be valid above $T_c$ and in the chiral limit $m\to 0$.
Being $\chi_\pi=\tilde\chi_{\eta,conn}$ and $\chi_\delta=\tilde\chi_{\sigma,conn}$, we obtain the following generalization of Eq.~\eqref{SU2rest}:
\begin{equation}\label{chitopN}
\chi_\pi-\chi_\delta = \tilde\chi_{\eta,conn}-\tilde\chi_{\sigma,conn} = \frac{2(N_f-1)}{N_f}\tilde\chi_{\sigma,disc} = -\frac{2(N_f-1)}{N_f}\tilde\chi_{\eta,disc} = 4(N_f-1)\lim_{m\to 0}\frac{\chi_{top}}{m^2}.
\end{equation}
Therefore, $\chi_\pi-\chi_\delta\propto\displaystyle\lim_{m\to 0}\chi_{top}/m^2$ is \textit{always} valid in the chirally restored phase. However, for $N_f>2$, if we use the above-mentioned DIGA prediction for $\chi_{top} = \mathcal{O}(m^{N_f})$ in the chiral limit $m\to 0$, for high temperatures $T$ \cite{GPY1981}, we conclude that $\chi_\pi-\chi_\delta\to0$ in the chiral limit at high temperatures: in this case, therefore, the $U(1)_A$ anomaly seems to have no effects on the chiral susceptibilities, in agreement with the statements of Refs. \cite{96Evans,96Lee,BCM1996}. Nevertheless, the $U(1)_A$ breaking could still manifest (and we shall argue that it \textit{does}) taking the expectation values of proper operators with a higher number ($2N_f$) of quark fields. We finally note that, as a result of the chiral susceptibilities being $U(1)_A$ symmetric, the condition \eqref{conditionN1} becomes $\tilde\chi_\sigma=\tilde\chi_{\sigma,conn}=\chi_\pi$, as in the case $N_f=2$.

\subsection{Zero-mode contribution}

In the chirally restored phase, the $U(1)_A$-breaking difference $\chi_\pi-\chi_\delta$ becomes proportional to the topological susceptibility $\chi_{top}$ [see Eqs.~\eqref{SU2rest} and \eqref{chitopN}], which receives a nonzero contribution only from the Dirac zero modes. We recall that these (discrete) modes are associated to the topological charge $Q$ through the Atiyah-Singer theorem \cite{84Atiyah} and that the topologically nontrivial gauge configurations are the ones that allow the $U(1)_A$ anomaly to have physical effects: therefore, it is natural to expect the Dirac zero modes to play a fundamental role in the $U(1)_A$ breaking. Indeed, these zero modes give rise to the following contribution in the spectral density, which explicitly breaks the $U(1)_A$ symmetry:
\begin{equation}\label{rho_0}
\rho(\lambda,m)\vert_0 = \frac{\delta(\lambda)}{V}\langle n_0[A]\rangle = \frac{\delta(\lambda)}{VZ}\int DA\,e^{-S_G[A]}[\det(\slashed{D}[A]+m)]^{N_f} n_0[A],
\end{equation}
where $n_0[A]$ is the number of zero modes of the gauge-field configuration $A_\mu$. Now, if the Dirac spectrum is \emph{discrete} (see the discussion below), the term $[\det(\slashed{D}[A]+m)]^{N_f}$ can be expressed as
\begin{equation}\label{det}
[\det(\slashed{D}[A]+m)]^{N_f} = \prod_k(-i\lambda_k[A] +m)^{N_f} = m^{N_fn_0[A]}\overline{\prod_k}(m^2+\lambda_k^2[A])^{N_f},
\end{equation}
where $\overline{\prod}_k$ is restricted to the nonzero modes. The Atiyah-Singer theorem \cite{84Atiyah} implies that $n_0=n_L+n_R\geq |Q|$, and in each topological sector (corresponding to a given $Q$) of the path integral one can always find a gauge-field configuration with $n_0=|Q|$, which will dominate in the chiral limit:
\begin{equation}\label{n0expression}
\langle n_0\rangle\underset{m\rightarrow 0}{\sim} \frac{1}{Z}\sum_{Q=-\infty}^{\infty}m^{N_f|Q|}|Q|\int DA_Q\,e^{-S_G[A_Q]}\overline{\prod_k}\lambda_k[A]^{2N_f}.
\end{equation}
The full path integral will then be dominated by the gauge configurations with $|Q|=1$ [i.e., the so-called \emph{instantons} ($Q=1$) or \emph{anti-instantons} ($Q=-1$)], resulting in a quark-mass dependence of the type $\langle n_0\rangle\sim m^{N_f}$. It is then convenient to express $\rho(\lambda,m)\vert_0$ as
\begin{equation}\label{deltaC}
\rho(\lambda,m)\vert_0 = \mathcal{C}m^{N_f}\delta(\lambda),\quad \textrm{with}\quad \mathcal{C}\equiv \frac{1}{V}\frac{\langle n_0\rangle}{m^{N_f}}.
\end{equation}
From the previous discussion, it follows that $\mathcal{C}$ is a nonzero constant in the chiral limit. Furthermore, from Eq.~\eqref{n0expression}, i.e., $\langle n_0\rangle\sim \langle |Q|\rangle$ as $m\to 0$, and since the dominant contribution comes from the topological sectors $Q=\pm1$, one can infer that
\begin{equation}
\langle n_0\rangle\sim \langle |Q|\rangle\sim \langle Q^2\rangle=V\chi_{top}
\end{equation}
in the chiral limit, so that we expect Eq.~\eqref{deltaC} to be nonvanishing in the thermodynamic limit as well.
Moreover, $\chi_{top} \sim \frac{\langle n_0\rangle}{V} = \mathcal{C} m^{N_f}$, in agreement with the above-mentioned DIGA prediction \cite{GPY1981}.

The problem with the above reasoning is that, by using Eq.~\eqref{det}, we have assumed a \emph{discrete} Dirac spectrum, while what we expect in the thermodynamic limit is a superposition of a \emph{continuous} spectrum and a \emph{discrete} one:\footnote{In some sense, by considering only the discrete spectrum we have taken the wrong order of limits: $m\to0$ \emph{before} $V\to\infty$. Actually, it has been argued that the order of limits no longer matters above the chiral transition \cite{96Evans}, since in this case there is no spontaneous-breaking phenomenon.} therefore, the quark-mass dependence could be more complicated. Still, gauge configurations such that the Dirac operator $\slashed{D}[A]$ has $n_0=1$ discrete zero mode give rise to a term $\sim m^{N_f}\delta(\lambda)$ in the spectral density.
Of course, this contribution must then be averaged in the full path integral: nevertheless, this contribution could survive the integration. Therefore, let us study the effects of the zero modes on the chiral susceptibilities, supposing that, near the chiral limit ($m\to 0$) and for $T>T_c$, the spectral density can be expressed as
\begin{equation}\label{rhodelta}
\rho(\lambda,m) = \rho(\lambda,m)\vert_0 + \ldots = \mathcal{C}m^{N_f}\delta(\lambda) + \ldots
\end{equation}
[This term was proposed in Ref. \cite{96Schafer} and further analysed in Refs. \cite{lat2012,lat2014,Ding2021}, where the authors found numerical evidences of the contribution \eqref{rhodelta} at high temperature.]
Substituting the expression \eqref{rhodelta} into Eq.~\eqref{rhoSigma} one obtains
\begin{equation}\label{rhodelta_Sigma}
\Sigma = N_fm\int_{-\infty}^{\infty}d\lambda\,\frac{\rho(\lambda,m)}{m^2+\lambda^2} = \mathcal{C}N_fm^{N_f-1} + \ldots,
\end{equation}
and then, from Eq. \eqref{chiralsuscN},
\begin{equation}\label{rhodelta_eta,conn}
\tilde\chi_{\eta,conn} = \frac{2}{N_f}\frac{\Sigma}{m} = 2\mathcal{C}m^{N_f-2} + \ldots,
\end{equation}
\begin{equation}\label{rhodelta_sigma,conn}
\tilde\chi_{\sigma,conn} = -2\int_{-\infty}^{\infty}d\lambda\,\rho(\lambda,m)\frac{m^2-\lambda^2}{(m^2+\lambda^2)^2} = -2\mathcal{C}m^{N_f-2} + \ldots,
\end{equation}
\begin{equation}\label{rhodelta_sigma,disc}
\tilde\chi_{\sigma,disc} = 2\int_{-\infty}^{\infty}d\lambda\,\frac{\partial\rho(\lambda,m)}{\partial m}\frac{m}{m^2+\lambda^2} = 2N_f\mathcal{C}m^{N_f-2} + \ldots,
\end{equation}
so that, in particular,
\begin{equation}\label{rhodelta_pi-delta}
\chi_{\pi}-\chi_{\delta} = \tilde\chi_{\eta,conn}-\tilde\chi_{\sigma,conn} = 4\mathcal{C}m^{N_f-2} + \ldots.
\end{equation}
Evidently, if $N_f=2$, the zero-mode contribution \eqref{rhodelta} results in a $U(1)_A$-broken scenario, since the $U(1)_A$ order parameter $\chi_\pi-\chi_\delta$ differs from zero in the chiral limit $m\to 0$.
Instead, the $SU(2)_V \otimes SU(2)_A$ chiral symmetry is restored, as it must be for $T>T_c$: in fact, from Eq. \eqref{rhodelta_Sigma} one sees that the chiral condensate $\Sigma$ vanishes in the chiral limit $m\to 0$, and, moreover, using Eqs. \eqref{chiralsuscN} and \eqref{rhodelta_eta,conn}--\eqref{rhodelta_sigma,disc} in the case $N_f=2$, one can immediately verify that the $\pi$ and $\sigma$ channels become degenerate ($\chi_\pi = \chi_\sigma$) and (using also the fact that $\chi_{top} = \mathcal{C} m^2 + \ldots$) that the $\delta$ and $\eta$ channels become degenerate, too ($\chi_\delta = \chi_\eta$).\\
On the other hand, if $N_f>2$, all the expressions \eqref{chiralsuscN} and \eqref{rhodelta_eta,conn}--\eqref{rhodelta_pi-delta} for the chiral susceptibilities vanish in the chiral limit $m\to 0$ (including $\tilde\chi_{\eta,disc} = -2N_f\frac{\chi_{top}}{m^2}$, being $\chi_{top} = \mathcal{C} m^{N_f} + \ldots$), in agreement with the well-known fact that only proper $n$-point quark-field correlation functions with $n\geq 2N_f$ can manifest the $U(1)_A$ breaking \cite{96Evans,96Lee,BCM1996}. In this case, $\chi_\pi-\chi_\delta$ no longer represents an order parameter for the $U(1)_A$ symmetry: in order to investigate the fate of this symmetry, one has to study functional averages of operators containing at least $2N_f$ quark fields. The $U(1)$ axial condensates, which we investigate in sections 2 and 3 of this paper, are precisely of this kind.

\newpage
	

\renewcommand{\Large}{\large}

\end{document}